\documentstyle[12pt,procsla,psfig]{article}

\catcode`\@=11
\long\def\@makefntext#1{
\protect\noindent \hbox to 3.2pt {\hskip-.9pt  
$^{{\ninerm\@thefnmark}}$\hfil}#1\hfill}              
  
\def\@makefnmark{\hbox to 0pt{$^{\@thefnmark}$\hss}}  
	
\def\ps@myheadings{\let\@mkboth\@gobbletwo
\def\@oddhead{\hbox{}
\rightmark\hfil\ninerm\thepage}   
\def\@oddfoot{}\def\@evenhead{\ninerm\thepage\hfil
\leftmark\hbox{}}\def\@evenfoot{}
\def\sectionmark##1{}\def\subsectionmark##1{}}

\newcommand{\lsim}{\mathrel{\mathop{\kern 0pt \rlap
  {\raise.2ex\hbox{$<$}}}
  \lower.9ex\hbox{\kern-.190em $\sim$}}}
\newcommand{\gsim}{\mathrel{\mathop{\kern 0pt \rlap
  {\raise.2ex\hbox{$>$}}}
  \lower.9ex\hbox{\kern-.190em $\sim$}}}
\newcommand{\beq}    {\begin{equation}}
\newcommand{\eeq}    {\end{equation}}
\newcommand{\beqarr} {\begin{eqnarray}}
\newcommand{\eeqarr} {\end{eqnarray}}
\newcommand{\barr}   {\begin{array}}
\newcommand{\earr}   {\end{array}}

\begin{document}


\title{Dark Matter and its Particle Candidates
\footnote{
Lectures given at the Fifth School on Non--Accelerator Particle
Astrophysics (Abdus Salam International Centre for Theoretical Physics,
Trieste, June--July 1998), to appear in the School Proceedings (Eds. 
R.A. Carrigan, Jr., G. Giacomelli and N. Paver)}
\setcounter{footnote}{3}
\footnote{Report no. DFTT 20/99, FTUV/99--24, IFIC/99--26}
}

\author{
\bf A. Bottino $^a$ and N. Fornengo $^b$ \\ 
\\
$^a$ Dipartimento di Fisica Teorica, Universit\`a di Torino and \\
     INFN, Sezione di Torino, via P. Giuria 1, I--10125 Torino, Italy \\
$^b$ Instituto de F\'\i sica Corpuscular -- C.S.I.C. \\
Departamento de F\'\i sica Te\`orica, Universitat de Val\`encia \\
c./ Dr. Moliner 50, E--46100 Burjassot, Val\`encia, Spain \\
{\sf E--mail: bottino@to.infn.it, fornengo@flamenco.ific.uv.es}
\vspace{6mm}
}

\date{ }

\maketitle

\begin{abstract}

In these lectures we first briefly review the main observational facts which 
imply that most part of matter in the Universe is not visible and
some recent intriguing experimental data which would point to a  significant
contribution to $\Omega$ due to a cosmological constant. We subsequently
discuss some particle candidates for dark matter, with particular emphasis for 
the neutralino. We present the main properties of this particle, also in the
light of the most recent experimental results in direct search for relic
particles; furthermore, we discuss the perspectives for their indirect searches. 

\end{abstract}  

\vspace{1cm}

\section{Dark Matter and Cosmology}

The two pillars of standard cosmology are the Einstein equations

\beq
{\rm R}_{\mu\nu} - {1\over 2} {\cal R} {\rm g}_{\mu\nu} = 
8 \pi {\rm G} {\rm T}_{\mu\nu} + {\rm \Lambda} {\rm g}_{\mu\nu} \, ,
\label{eq:einstein}
\eeq

\noindent
(where ${\rm R}_{\mu \nu}$ is the curvature (Ricci) tensor, $\cal R$ is its 
trace, ${\rm T}_{\mu \nu}$ is the stress-energy tensor, ${\rm \Lambda}$ the 
cosmological constant) and the Robertson-Walker metric 

\beq
ds^2 = dt^2 - {\rm R}^2(t) \left({ \frac{dr^2}{1-k r^2} + r^2 d{\theta}^2 + 
r^2 \sin^2 \theta \, d{\phi}^2} \right) \, ,
\label{eq:rw}
\eeq

\noindent
where $R = R(t)$ is the cosmic scale factor and $t,r,\theta,\phi$ are the 
comoving coordinates \cite{weinberg,kt,peebles,roos,kolb}. 

We recall that Eq. (\ref{eq:rw})
 follows from the assumption that the distribution of matter 
and radiation in the Universe is isotropic and homogeneous.  
 When the 
coordinates are appropriately rescaled, the values 
$k = +1, 0, -1$ define the space curvature to be positive, zero 
and negative, respectively. 
Useful standard quantities for the description of the expanding 
Universe are the Hubble parameter (expansion rate of the Universe) 
$H = \dot{\rm R}/R$ and the  deceleration parameter 
$q = - \ddot{\rm R} \cdot {\rm R}$/$\dot{\rm R}^2$, 
whose values at present epoch are denoted as ${\rm H}_0$ and ${\rm q}_0$.

By combining Eqs. (\ref{eq:einstein}) -- (\ref{eq:rw}), one  obtains the 
Friedmann equation 

\beq
{\rm \Omega}_m + {\rm \Omega}_{\rm \Lambda} - k/({\rm H}^2 {\rm R}^2) = 1\, ,  
\label{eq:friedmann}
\eeq

\noindent
where ${\rm \Omega}_m$ is the ratio of the average matter--energy density $\rho$ 
to the critical density $\rho_c = 3 {\rm H}^2/(8 \pi G)$, 
$\Omega_m = \rho/\rho_c$, and ${\rm \Omega}_{\rm \Lambda} = 
{\rm \Lambda} /(3 {\rm H}^2)$. The critical density is given by 
$\rho_c =  1.9 \cdot 10^{-29}\, h^2$  g\ cm$^{-3}$, when the Hubble 
constant is parametrised as follows:
$h = {\rm H}_0/(100$ km\  s$^{-1}$ Mpc$^{-1}$). 

{}From Eqs. (\ref{eq:einstein}) -- (\ref{eq:rw}) it also follows that at 
present time, when the radiation contribution to $\Omega$ can be set
to zero, the value of the deceleration parameter is given by 

\beq
q_0 = \frac{1} {2} {\Omega}_m - {\Omega}_{\Lambda}\, . 
\label{eq:q}
\eeq

The features of the evolution of our Universe depend on the actual 
values to be assigned to the cosmological parameters previously 
defined. In what follows we briefly summarize some of the main 
observational  data about these parameters.

\section{Age of the Universe and Expansion Rate}

The evaluation of 
the present age of the Universe, $t_0$, depends on the expansion rate 
and on the various components of $\Omega$; 
therefore measurements  of $t_0$ and 
of the Hubble constant provide information on the size of 
$\Omega_m$ and $\Omega_{\Lambda}$ (see, for instance, Refs. 
\cite{kt,freedman}).

A recent determination of $t_0$ provides the value: 
$t_0 = 11.5 \pm 1.3$ Gyr \cite{cdkk}, with a $95\%$ C.L. lower 
bound of 9.5 Gyr. Recent measurements of the Hubble constant by the Hubble
Space Telescope Key project 
give the following range \cite{freedman,madore}: 
${\rm H}_0 = 73 \pm 6 (stat) \pm 8 (syst)$ km s$^{-1}$ Mpc$^{-1}$. 
In view of the still persisting sizeable spread in the $h$ values, due to a host 
of independent measurements (for a review, see for instance 
\cite{freedman})  
in the following we will, conservatively, consider a rather wide range:
$0.5 \leq h \leq 0.9$.    

If, for sake of illustration, we take $h \simeq 0.7$, 
it is easy to show  that $t_0 \sim 11.5$ Gyr would require either  
$\Omega_m \lsim 0.3$ or $\Omega_m \sim 0.5$, according to whether or not 
we allow for a non-vanishing cosmological constant, such that 
$\Omega_{\Lambda} = 1 - \Omega_m$.

\section{Observational Evidence for Dark Matter}

 The parameter $\Omega_m$ may be derived from astronomical 
determinations of the 
average mass-to-light ratio $M/L$ for various astrophysical objects 
(see, for instance, Ref. \cite{sap} for an updated review)  

\beq
 \Omega_m = \frac{\rm M}{\rm L} \frac{\it L}{\rho_c}\, ,
\label{eq:ml}
\eeq

\noindent
where $L$ is the luminosity density: 
${\it L} = 1.6 \cdot 10^8\, h\, {\rm L}_{\odot}$ Mpc$^{-3}$. Using 
$\rho_c = 2.8 \cdot 10^{11}\, h^2\, {\rm M}_{\odot}$ Mpc$^{-3}$, one gets 

\beq
 \Omega_m = \frac{h^{-1}}{1500}\, \frac{{\rm M}/{\rm L}}
{{\rm M}_{\odot}/{\rm L}_{\odot}} \, . 
\label{eq:aa}
\eeq

{}From the $M/L$ ratio measured in stars, 
$M/L \sim (3-9) {\rm M}_{\odot}/{\rm L}_{\odot}$  we obtain

\beq
  0.002 \;h^{-1} \leq \Omega_{vis} \leq 0.006 \; h^{-1} \, .  
\label{eq:bb}
\eeq

\subsection{Rotational curves of spiral galaxies}

Presence of dark matter in single galaxies is apparent from the 
flatness of the plot of the rotational velocities versus the 
galactocentric radius, well beyond the distribution of the visible 
matter. {}From these measurements one derives 
$M/L \simeq 70  \, {\rm M}_{\odot}/{\rm L}_{\odot} (R_{halo}/100$ kpc). 
Then we obtain 

\beq
   \Omega_{halo} \sim 0.05\, h\, R_{halo}/100\, \mbox{{\rm kpc}}\, , 
\eeq

\noindent
a result which, compared to the range (\ref{eq:bb}) for $\Omega_{vis}$, is 
indicative of the presence of dark matter at the 
level of single galaxies. 

\subsection{Clusters of galaxies}
 
Existence of dark matter at the level of clusters of galaxies may be 
established by a number of methods: X-ray emission by hot gas in 
intra cluster plasma, measurements of velocity dispersion and 
strong gravitational lensing. 

We report here some results derived from 
 measurements of the X-ray emission \cite{wf}. For rich clusters 
of galaxies one finds 
$M/L = (300 \pm 100)\,h \,{\rm M}_{\odot}/{\rm L}_{\odot}$, which 
gives 

\beq
   \Omega_{cluster} \simeq 0.2 \pm 0.07 \, . 
\label{eq:cc}
\eeq

\noindent
The baryonic content $\rm \Omega_b$ is established to be: $\sim 6\;h^{-3/2}\%$ 
for gas and $ \gsim 4\%$ for stars; then 

\beq
  \frac{\Omega_b}{\Omega_{cluster}} \gsim 0.04 + 0.06 h^{-3/2} \, .  
\label{eq:dd}
\eeq

On the other hand, the Big Bang nucleosynthesis provides the following 
estimate for $\Omega_b$ \cite{sap,nucl}: 
\begin{equation}
0.009 \lsim  \Omega_b  h^2 \lsim 0.02 \, ,
\label{eq:ob}
\end{equation}
or, taking $0.5 \leq h \leq 0.9$, 
$0.01 \lsim {\rm \Omega_b}  \lsim 0.08$.
Combining this result with Eq.(\ref{eq:dd}) we obtain 

\beq
{\Omega}_{cluster} \lsim 0.2 - 0.4 \, .
\label{eq:ee}
\eeq

\subsection{Large-scale Velocity Flows}

    Distribution  of matter over large scales may be inferred from the 
peculiar motion of the gravitationally--induced inflow. 
Let us consider the case of the Local Supercluster, centered near the 
Virgo cluster; we stay on the edge of this cluster, at a distance of 
R = 12 $h^{-1}$ Mpc. The radial inward peculiar velocity averaged over 
the surface is given by \cite{peebles2} 

\beq
  \bar v = \frac{1}{3} {\rm H}_0 \, {\rm R} \delta_m {\Omega}^{0.6} \, . 
\label{eq:xx}
\eeq

\noindent
where ${\Omega}^{0.6}$ represents the factor due to expansion and 
$\delta_m$ denotes the mass contrast, related to the contrast in 
galaxy counts ($\delta_g$)  by the relation 
$\delta_m = \delta_g/b$, $b$ being the 
bias factor. Using the observational values \cite{peebles2}:
$\bar v = 200 \pm 25$ km s$^{-1}$, $\delta_g = 2.3 \pm 0.7$, 
and taking $b \simeq 1$, one obtains 

\beq
    {\Omega}_m \simeq 0.1_{-0.05}^{+0.1}\, . 
\eeq

\subsection{A few first conclusions}

{}From the previous evaluations for $\Omega_{vis}$ and  
$\Omega_m$, and from the range for $\rm \Omega_b$ in Eq.(\ref{eq:ob})  we derive 
the following important points: 

\begin{itemize}

\item a large amount of matter in the Universe is not visible;

\item some of this dark matter is  baryonic;

\item a significant amount of dark matter is \underline{non}-- baryonic.

\end{itemize}

As usual we divide particle dark matter into two categories: 
Hot Dark Matter (HDM) and Cold Dark Matter (CDM), according to whether 
the particles are relativistic or non-relativistic at their decoupling 
from the primeval plasma.

\section{Singling out different contributions to $\rm \Omega$}

We briefly report now some important observational results and analyses 
which provide further clues toward a determination of the various components 
to $\rm \Omega$. The two main issues are: what, if any, is the size  
of $\rm \Omega_{\Lambda}$ and what is $\Omega_m$ made of?

\subsection{Formation of cosmological structures}

A standard method for testing different dark matter models is to 
compare the power spectrum of the density fluctuations $P(k)$ 
 with measurements of the Cosmic Microwave Background 
 Radiation (CMBR) anisotropy  and measurements of the 
two-point correlation function in galaxy surveys 
\cite{kt,peebles,roos,kolb}. 
We recall that $P(k)$ is the Fourier transform of the correlation 
function between the density contrasts at two different points in 
space. It is customary to assume for $P(k)$ 
a power law, i.e. 
$P(k) \propto k^n$. The Harrison--Zel'dovich spectrum corresponds 
to $n = 1$. In the past, typically, the best performing cosmological 
models turned out to fall into the following categories: 

\begin{itemize}

\item HCDM model characterized by $\rm \Omega = 1$ and the following 
      composition: 
      $\rm \Omega_b \simeq 0.1$, $\rm \Omega_{\nu} \simeq 0.2$, 
      $\rm \Omega_{CDM} \simeq 0.7$; $h \simeq 0.5$, where 
      $\rm \Omega_b$, $\rm \Omega_{\nu}$ and $\rm \Omega_{CDM}$ are
      the contribution due to baryons, neutrinos (as HDM particles) 
      and to CDM particles, respectively. 

\item $\rm \Lambda$CDM model with $\rm \Omega = 1$ and 
      $\rm \Omega_{CDM} \simeq 0.3$, 
      $\rm \Omega_{\rm \Lambda} \simeq 0.7$; $h \simeq $ 0.7 -- 0.8.

\item TCDM model ($\equiv$ tilted CDM model) with a power spectrum 
      $P(k) \propto k^{0.8}$ and $\rm \Omega = 1 = \rm \Omega_{CDM} = 1$; 
      $h \simeq 0.5$. 

\end{itemize}

\subsection{Evidence for $\rm \Omega_{\rm \Lambda} \neq 0$ ?}

Recent measurements of high-redshift supernovae of type Ia 
\cite{perl1,perl2,reiss} point to 
an important contribution to 
$\rm \Omega$ due to $\rm \Lambda$, with a relatively small contribution from
$\Omega_m$. 
These data appear to be complementary to those derived from 
measurements of the CMBR \cite{new}. The joint use of these two sets of data, 
together with some other observational data, 
singles out the following ranges: 
$\Omega_m = 0.24 \pm 0.10$ and $\Omega_{\Lambda} = 0.62 \pm 0.16$
\cite{lineweaver}. 
These data and analyses are of the utmost interest for their 
potential implications, although a number of points need further 
clarification \cite{peebles}. 

One further hint for a rather low value of $\rm \Omega_m$ is 
provided by the time evolution of the number density of clusters. 
In Ref.\cite{neta} observational data on cluster abundance in the 
redshift range $0 \lsim z \lsim 1$ is used to derive the estimate 
$\rm \Omega_m \simeq 0.2^{+0.15}_{-0.1}$. 

Though a cautionary attitude is in order here, it is important to 
remark that all these different ways of determining 
$\rm \Omega_m$ point  to a relative small value for 
this quantity: $\rm \Omega_m \lsim$ 0.3 -- 0.4.
This feature, if confirmed by further observational data, has 
profound implications for the phenomenology related to DM 
particle candidates, as will be illustrated in the following.

\section{Particle candidates for dark matter}

\subsection{Baryons}

As was already noticed in Sect. III, some contribution to DM is 
provided by baryons. This conclusion is drawn from the fact 
that the Big Bang nucleosynthesis provides the estimate 
$0.009 \lsim  \Omega_b  h^2 \lsim 0.02$, which, together with 
Eq. (\ref{eq:bb}),  implies ${\rm \Omega_b} > {\rm \Omega_{vis}}$. 

Direct search for non-luminous baryonic dark matter, under 
the form of microlensing objects, has been undertaken 
since the seminal paper of Ref.\cite{pac}. Recent results are reviewed 
in \cite{sap,spiro,stubbs}. The main features of the data may be summarized 
as follows. The EROS Collaboration \cite{eros} excludes that microlensing
objects of masses in the range  $(5\cdot10^{-8} - 10^{-2})\, M_{\odot}$ may
make up more than 20\% of the halo density, whereas the 
 MACHO Collaboration \cite{macho} delimits a likelihood contour (at 
95\% C.L.) for masses $\sim (10^{-1} - 1) \, M_{\odot}$ with a best-fit 
value for the halo fraction around 0.5. For further details see, for instance, 
Refs. \cite{sap,spiro,stubbs}. 
For a critical view of the microlensing events, see Ref.\cite{FFG}.
An interesting scenario related to
baryonic dark matter is the one depicted
in Ref.\cite{jetzer}.

\subsection{Non-baryonic DM candidates}

Particle physics offers a large variety of particles, which would 
have decoupled from the primeval plasma at the time (freeze--out time), 
when the 
interaction rate became smaller than the cosmic expansion rate;
these particles would then be floating around in our 
Universe as relics.  

These fossil particles would or would not significantly 
contribute to the average cosmic density depending on their actual 
number density and mass. The most obvious example is provided by 
light fossil neutrinos, whose relic abundance may be easily 
evaluated (see, for instance Ref.\cite{kt}) and turns out to be 

\beq
{\rm \Omega}_{\nu} h^2 = \frac {\sum m_{\nu}} {93 \, \mbox{\rm eV}}\, ,
\label{eq:neu}
\eeq

\noindent
where the sum is over the neutrino flavours (for each flavour, 
neutrino and antineutrino are counted together). 
Therefore 
the relevance of these fossil neutrinos for the Universe matter 
density depends on whether  their mass is of order of a few 
eV or much smaller (provided neutrinos are massive at all). 

Possible indications for non--vanishing neutrino masses are from: 
1) solar neutrinos \cite{solar,smy}, 2) atmospheric neutrinos 
\cite{sk,macro,habig}, and 3) the LSND experiment \cite{lsnd}. 
All these data refer to oscillation measurements, and then are 
not sensitive to individual neutrino masses, but only to 
differences in their squares.  Typically, atmospheric neutrino 
experiments give 
$\Delta m^2 \simeq (2 \div 6) \times 10^{-3} \mbox{{eV}}^2$; 
solar neutrino experiments 
can be explained either by a $\Delta m^2 \simeq 10^{-5} \mbox{{eV}}^2$, 
in case of matter--enhanced oscillations, or by a 
$\Delta m^2 \simeq 10^{-10} \mbox{{eV}}^2$, in case of vacuum oscillations; 
 LSND data suggest 
$0.2 \,\mbox{{eV}}^2 \lsim \Delta m^2  \lsim 2 \,\mbox{{eV}}^2$. 
 These results may be 
compatible with a sizeable value of relic abundance, 
$\rm \Omega_{\nu} \sim 0.2$, such as preferred by some calculations of 
cosmological structures. However, if taken at their face values, they only 
imply ${\rm \Omega_{\nu}} \gsim 0.02 \, (0.5/h)^2$ \cite {prim}.

A different candidate is the axion, whose motivation in particle physics 
is related to the strong CP-problem in QCD \cite{peccei}. A discussion 
of this candidate is beyond the scope of the present report; for a 
comprehensive review on its possible cosmological relevance and on the 
experimental efforts for detecting it as a relic particle, see for instance 
Ref. \cite{axion}.  

Among the particle candidates for CDM the most favorite one is certainly 
the Lightest Supersymmetric Particle (LSP), under the condition that it is 
weakly interacting. This candidate is discussed in the 
following section.

\section{Supersymmetric dark matter particles}

In order to be a dark matter candidate a particle has to be weakly
interacting and stable (or, at least, long lived on cosmological time--scales). 
A very interesting perspective for such a candidate is offered, in the framework
of 
supersymmetric theories with R-parity conservation, by the lightest susy 
particle (provided this is indeed weakly interacting). 
Different candidates have been considered: the neutralino \cite{neutralino}
or the sneutrino\cite{snu} in gravity mediated models, the gravitino\cite{gravitino} 
or messenger fields\cite{messenger} in gauge mediated theories, the axino\cite{axino}, 
stable non--topological solitons (Q--balls)\cite{qballs},
heavy non--thermal relics\cite{zillas} or others.

Among the different candidates, the most promising one turns out to be
the neutralino, since its relic abundance may be sizeable,
at the level required to explain the CDM content of the 
Universe and, at the same time, its detection rates 
may be accessible to experimental searches of different kinds. 

The phenomenology of neutralino dark matter has been studied extensively
in the Minimal Supersymmetric extension of the Standard Model (MSSM)
\cite{susy}.
This model incorporates the same gauge group as the Standard Model
and the supersymmetric extension of its particle content. The
Higgs sector is slightly modified as compared to that of the Standard
Model: the MSSM requires
two Higgs doublets $H_1$ and $H_2$ in order to give mass both to down-- and
up--type quarks and to cancel anomalies. After electroweak symmetry
breaking,
the physical Higgs fields consist of two
charged particles and three neutral ones: two scalar fields ($h$ and $H$) and
one pseudoscalar ($A$). The Higgs sector is specified at the tree level by
two independent parameters:
the mass of one of the physical Higgs fields and the ratio of the two vacuum
expectation values, usually defined as $\tan\beta =\, <H_2> / <H_1>$.
The supersymmetric sector of the model introduces some other free parameters:
the mass parameters $M_1$, $M_2$ and $M_3$ for the supersymmetric partners
of gauge fields (gauginos), the Higgs--mixing parameter $\mu$ and, in general,
all the masses of the scalar partners of the fermions 
(sfermions) and all the trilinear couplings which enter in the 
superpotential.
In the MSSM it is generally assumed that the gaugino masses are related
by expressions induced by grand--unification. Specifically, the two parameters 
which are relevant for neutralino phenomenology are linked by: 
$M_1= (5/3) \tan^2 \theta_W M_2$. The other usual assumptions are that
all slepton mass parameters are taken as degenerate to a common
value $m_0$ and that all the trilinear couplings are vanishing
except the ones of the third family which are set to a common
value $A$. In summary, the free parameters of the model
are six: $M_2$, $\mu$, $\tan\beta$, $m_A$, $m_0$ and $A$.

The neutralinos are four mass--eigenstates defined as 
linear superpositions of the
two neutral gauginos ($\tilde \gamma$ and $\tilde Z$) and the two
neutral higgsinos ($\tilde H_1$ and $\tilde H_2$)
\begin{equation}
\chi = a_1 \tilde \gamma + a_2 \tilde Z + a_3 \tilde H_1 + a_4 \tilde H_2\;.
\end{equation}
\noindent
The lowest--mass eigenstate may play the role of the lightest supersymmetric
particle in the MSSM, and may then constitute the dark matter candidate
in this model. It will be called the neutralino tout--court and its mass
denoted by $m_\chi$. 
To classify the nature of the neutralino it is useful to define a parameter 
$P \equiv a_1^2 + a_2^2$; the neutralino is called a {\it gaugino}, 
when $P > 0.9$, a {\it higgsino} when $P < 0.1$ and
{\it mixed} when $0.1 \leq P \leq 0.9$. 

The low--energy MSSM scheme is a phenomenological approach, whose
basic idea is to impose as few model--dependent restrictions as possible.
At a more fundamental level, it is natural to implement this
phenomenological scheme within the supergravity (SUGRA) framework
\cite{sugra}. One
attractive feature of the ensuing model is the connection between
soft supersymmetry breaking and electroweak
symmetry breaking, which would then be induced radiatively.
Usually, the low--energy phenomenology of SUGRA theories
constitutes a subset of the susy configurations which are
considered in the MSSM. A typical characteristic of
SUGRA models is in fact the presence of relatively strong 
correlations among the low--energy parameters, correlation which 
is absent in the MSSM.
In this report we will discuss neutralino dark matter phenomenology
in the less constrained MSSM model. Results for SUGRA schemes
can be found in the papers listed in Ref.\cite{sugra_dm}

\subsection{Neutralino relic abundance}

Neutralinos decouple from the hot plasma in the early
Universe when they are no longer relativistic. Their
present abundance is calculated by solving the Boltzmann
equation for the evolution of the density of particle
species\cite{kt} and can be written as:
\begin{equation}
\Omega_\chi h^2 = {\cal C} \, 
\frac{g_*^{1/2}(T_f)}{{g_*}_S (T_f)} \,
\frac{1}{\langle \sigma_{\mbox{\rm\scriptsize ann}} v_r 
\rangle_{\mbox{\rm\scriptsize int}}} \,
\end{equation}
where
\begin{equation}
{\cal C} = \frac{s_0}{0.264\, \rho_c\, M_P} = 8.7 \cdot 10^{-11} \,  
 \mbox{\rm GeV}^{-2} \, .
\end{equation}

In the previous Eqs., $g_*(T_f)$ and ${g_*}_S(T_f)$ denote the 
effective number of degrees of freedom for the energy
density and for the entropy density, respectively, 
evaluated at the freeze--out temperature $T_f$;
$\langle \sigma_{\mbox{\rm\scriptsize ann}} v_r 
\rangle_{\mbox{\rm\scriptsize int}}$ is the 
neutralino pair annihilation times the pair
relative velocity, averaged over the neutralino
thermal density distribution, integrated from the
freeze--out temperature down to the present temperature;
$s_0$ denotes the present entropy density,
$\rho_c$ is the critical density and $M_P$ is the Planck mass.

The critical quantity to be evaluated is the neutralino
annihilation cross section, which, depending on the
neutralino mass, can get contributions from the following
final states :
i) fermion--antifermion pair, ii) pair of neutral Higgs bosons,
iii) pair of charged Higgs bosons, 
iv) one Higgs boson-one gauge boson, v) pair of gauge bosons. 
The diagrams contributing to the final state i) are
Higgs--exchange Z--exchange diagrams in the s--channel, 
sfermion--exchange diagrams in the t--channel.
For the other final states, the contributions come from
Higgs and Z--diagrams in the s--channel, and either 
neutralinos or charginos exchange in the t--channel, 
depending on the electric charges of the final particles
\cite{noi_omega,omegah2}.
When the mass of the neutralino is close to the mass of
another supersymmetric particle, the process of
co--annihilation \cite{coann,suede_coann} can be present. 
In this case, the calculation of the annihilation cross 
section and of its thermal average has to take into account 
a large number competing interactions among the neutralino and 
its close--mass particles. In some  special instances the relic
abundance may be affected by co--annihilation effects by a 
sizeable amount\cite{suede_coann}.

In Fig. 1 we show $\Omega_\chi h^2$ as a function of the
neutralino mass $m_\chi$ \cite{noi_omega}.
We present here, as well as in the following, all the results 
in terms of scatter plots
which are obtained by varying the supersymmetric parameters
inside wide ranges.
Naturally, the parameter space is constrained by experimental
limits on Higgs bosons and supersymmetric particles searches
(for recent updates, see \cite{LEP}) and by measurements 
of rare processes, whose theoretical predictions are quite
sensitive to supersymmetric contributions. At present,
the most important is the decay 
$b \rightarrow s + \gamma$\cite{bsg_exp,bsg_theo}.

The scatter plot of Fig. 1 is obtained by varying 
the supersymmetric parameters in the following ranges:
$20\;\mbox{GeV} \leq M_2 \leq  500\;\mbox{GeV},\; 
20\;\mbox{GeV} \leq |\mu| \leq  500\;\mbox{GeV},\;
80\;\mbox{GeV} \leq m_A \leq  1000\;\mbox{GeV},\; 
100\;\mbox{GeV} \leq m_0 \leq  1000\;\mbox{GeV},\;
-3 \leq {\rm A} \leq +3,\;
1 \leq \tan \beta \leq 50$. 

The figure shows only those configurations which provide 
a value for the relic abundance lower than a cosmological 
upper bound which has been conservatively set as
$\Omega_\chi h^2 \leq 0.7$. The susy configurations 
which entail larger values of $\Omega_\chi h^2$ are
excluded by the lower limit on the age of the Universe 
\cite{cdkk}.  This constraint  is rather restrictive
on the susy parameter space.

\subsection{Detection of relic neutralinos}

Relic neutralinos would act as CDM during the process
of galaxy formation. It is therefore conceivable
that they may constitute all or part 
of the dark matter halo. These neutralinos
would be clustered in the Galaxy and hence 
possess a matter density distribution and a 
velocity distribution which depend on the 
dynamics of the Galaxy formation and evolution\cite{BT}.

Many different models have been discussed in the
literature for the {\em matter density distribution}
$\rho(\vec r)$ (see for instance Refs. \cite{nfw1,nfw2,kkbp}). 
This field is in rapid expansion, due to the high resolution simulations 
now at hand to investigate the structure of single galaxies \cite{moore}. 
In particular, these studies are expected to 
pin down the nature of the cuspy behaviour which appears to occur 
near the galactic center. Another very
interesting possibility which is currently  being investigated 
is that the halo could present a clumpy
distribution of dark matter together
with or in alternative to a smooth distribution.
The uncertainties in the shape profile, combined with
experimental uncertainties and the possibility
that some fraction of the dark halo is made of
baryonic dark matter in the form of MACHOS,
implies that the local value of the matter density
$\rho_l = \rho(\vec r_\odot)$ is somewhat uncertain.
At present, its most conservative range of variability
can be set as \cite{turner_rhol}:
0.1  GeV cm$^{-3}$ $\leq \rho_l \leq 0.7$ GeV cm$^{-3}$. 

The quantity $\rho_l$ refers to the total dark matter
density of the galactic halo. the neutralino local
density is, in general, a fraction of it, i.e.
$\rho_\chi = \xi \rho_l$, with $\xi \leq 1$. 
No exact  way to evaluate the
quantity $\xi$ is currently available.
A usual procedure is to consider the galactic halo
as composed entirely of neutralinos if, on the
average in the Universe, neutralinos are abundant
enough to explain all the CDM which is observed
at least on the galactic scale. This happens when
$\Omega_\chi h^2$ is larger than a value
$(\Omega h^2)_{\rm min}$ derived from astrophysical
observations. In this case it is possible to
set $\xi = 1$. If, on the contrary, 
$\Omega_\chi h^2 < (\Omega h^2)_{\rm min}$, neutralinos
are not able to explain all the CDM, even the
one which is needed at the galactic scale,
and therefore also locally in our Galaxy we expect them to
give only a fractional contribution $\xi$ to
$\rho_l$. In this case, the simplest choice is to set: 
$\xi = \Omega_\chi h^2 / (\Omega h^2)_{\rm min} $.
The quantity $ (\Omega h^2)_{\rm min}$ is estimated
to lie in the range
$0.01 \lsim (\Omega h^2)_{\rm min} \leq 0.2$.

The {\em velocity distribution} $f(\vec v)$ of dark matter is
usually assumed to be a Maxwellian distribution \cite{BT} 
(as seen from the galactic rest frame), with a 
velocity dispersion
$v_{\rm rms}$ which is directly related to the
asymptotic flat rotational velocity $v_\infty$ as:
$v_{\rm rms} = \sqrt{3/2}v_\infty$. In our Galaxy, 
the rotational velocity appears to be already flat
at the local position, and therefore we set 
$v_{\infty} = v_{\odot}$. The experimental determination
of the local rotational velocity is: 
$v_{\odot} = 230 \pm 50$ Km s$^{-1}$ 
(90 \% C.L.)\cite{koch}. 
The distribution is also truncated by an escape velocity
$v_{\rm esc}$ which falls in the range:
$v_{\rm esc} = 450 \div 650$ Km s$^{-1}$ (90 \% C.L.)\cite{leonard,cud}.
Modifications to the standard Maxwell--Boltzmann form have been examined
\cite{BT,evans}, but the  problem of determining  the correct 
form of the distribution of the  velocities in the
halo has  no clear  and simple   solution at  present,  both  
theoretically and observationally. Also the possibility that
the halo could possess a bulk rotation has been considered in the
literature\cite{fv_rot}.

Due to the possibility that neutralinos are
present in the halo of our Galaxy,
it is of great interest, both
for astrophysics and particle physics, to search for 
techniques capable of detecting  these 
dark halo particles. Several methods have been proposed
and in the following we will briefly review the ones
which, at present, appear to be more promising.

\subsection{Direct detection of relic neutralinos}

The most direct possibility to detect dark matter particles is
to look for their scattering with the nuclei of
a low--background detector\cite{witten}. The interaction of slow
halo neutralinos of mass $m_\chi \gsim 25$ GeV  with a detector
produces the recoil of a nucleus with energy $E_R$ of the order of
few to tens keV. The recoil energy can be measured by means of
various experimental techniques with different nuclear species,
like Ge, NaI, Xe, CaF$_2$, TeO$_2$\cite{taup_direct}.
The relevant quantity to be calculated and compared
with the experimental measurements is the differential detection rate
\begin{equation}
S_0(E_R) \equiv
\frac {dR}{dE_R}=N_{T}\frac{\rho_{\chi}}{m_{\chi}}
                    \int \,d \vec{v}\,f(\vec v)\,v
                    \frac{d\sigma}{dE_{R}}(v,E_{R}) \label{rate}
\label{eq:diffrate0}
\end{equation}
where $N_T$ is the number of the target nuclei per unit of mass,
$\rho_\chi$ is the local neutralino matter density,
$\vec v$ and $f(\vec v)$ denote the neutralino
velocity and velocity distribution function in the Earth frame
($v = |\vec v|$). The nuclear recoil energy is given by
$E_R={{m_{\rm red}^2}}v^2(1-\cos \theta^*)/{m_N}$,
where $\theta^*$ is the scattering
angle in the neutralino--nucleus center--of--mass frame,
$m_N$ is the nuclear mass and $m_{\rm red}$ is the neutralino--nucleus
reduced mass. Finally, $d\sigma/dE_R$ is the neutralino--nucleus 
differential cross section, which has a coherent contribution
due to Higgs-- and squark--exchange, and a spin dependent one
which originates from the exchange of the Z boson and squarks.
The coherent cross section is usually largely dominant over the
spin--dependent one.

Eq.(\ref{eq:diffrate0}) refers to the situation of a monoatomic
detector, like the Ge detectors. For more general situations, like
for instance the case of NaI, the generalization is straightforward.
{}From those experimental data on the nuclear recoil spectrum which do not 
provide a signal--to--background discrimination, upper limits to the 
neutralino--nucleus cross section as a function of the
neutralino mass may be set by employing Eq.(\ref{eq:diffrate0}) \cite{bdmsbi}.
In the case of coherent interaction, Fig. 2
shows, as a solid line, the present most stringent upper 
limit\cite{DAMA_uplim} on the
quantity $\xi \sigma^{(\rm nucleon)}_{\rm scalar}$ vs.
$m_\chi$, where $\sigma^{(\rm nucleon)}_{\rm scalar}$ denotes
the scalar elastic cross section of a neutralino off a nucleon.
The astrophysical parameters are chosen as: 
$\rho_l = 0.3$ GeV cm$^{-3}$, $v_0 = 220$ Km s$^{-1}$,
$v_{\rm esc} = 550$ Km s$^{-1}$ and
$ (\Omega h^2)_{\rm min} = 0.01$.

In this plot, we also show a scatter plot of
the same quantity calculated in the MSSM\cite{noi_diretta,diretta_theo}. 
The susy configurations 
have been varied in the ranges quoted in Sect. 6.1.
We stress that the cosmological bound $\Omega_\chi h^2 \leq 0.7$
has been applied \cite{note_damour}.

In the case of direct detection, a typical signature consists 
in the annual modulation of the detection rate\cite{ann_mod_th}.
During the orbital motion of the Earth around the Sun,
the change of direction of the relic particle velocities
with respect to the detector induces a time dependence in the 
differential detection rate, i.e.
$S(E_R,t) = S_0 (E_R) + S_m (E_R) \cos [\omega (t-t_0)]$,
where $\omega = 2\pi/365$ days and $t_0 = 153$ days
(June 2$^{\rm nd}$). $S_0 (E_r)$ is the average
(unmodulated) differential rate defined in 
Eq.(\ref{eq:diffrate0})and $S_m (E_R)$ is the
modulation amplitude of the rate.
The relative importance of $S_m (E_R)$ with respect to $S_0 (E_R)$ 
for a given detector, depends both on the mass of the dark matter 
particle and on the value of the recoil energy where the effect is looked 
at. Typical values of $S_m (E_R)/S_0 (E_R)$
for a NaI detector range from a few percent up to $\sim$ 15\%,
for neutralino masses of the order of 20--80 GeV and recoil energies
below 8--10 KeV.

The search for the annual modulation effect is currently undertaken by
the DAMA/ NaI Collaboration\cite{DAMA_longrep}. 
The analysis of their data over two
years of data--taking 
provides the indication of a modulated
signal \cite{DAMA}.  This result
can be shown as the closed contour in the 
$\xi \sigma^{(\rm nucleon)}_{\rm scalar}$ vs. $m_\chi$ 
plane displayed  in Fig. 2. The region inside the contour is compatible
with a modulation signal at 2--$\sigma$ level. 
The contour takes into account also the uncertainties in 
astrophysical velocities, as discussed in \cite{bellietal}.
Fig. 2 shows that there exist susy configurations which would be able to
explain such an effect. In the papers of
Refs.\cite{bellietal,noi_DAMA}
it has been proved that some of these configurations are 
explorable at accelerators and/or by WIMP indirect experiments and that the 
relevant relic neutralinos might behave as major components of cold dark 
matter. For an analysis of these configurations in a SUGRA scheme, see
also Ref.\cite{an}.

\subsection{Indirect detection: neutrino fluxes from Earth and Sun}

Other ways of detecting dark matter neutralinos
rely on the possibility to detect the products of neutralino annihilations. 
One perspective is to observe a neutrino signal coming from celestial
bodies, namely Earth and Sun, where the neutralinos may have been
captured and accumulated during the lifetime of the macroscopic
body. The sequence of the physical processes which could produce these
signals are: i) capture of relic neutralinos by the macroscopic 
bodies; ii) subsequent  accumulation of these
particles in a region around the centre of these celestial objects;
iii) pair annihilation of the accumulated neutralinos which would
generate, by decay of the particles produced in the  various
annihilation final states, an output of high--energy neutrinos.
Since the process of annihilation takes place inside a medium
(i.e., the interior of the Earth or the Sun), the annihilation
process is perturbed as compared to the annihilation in 
vacuum. This effect can be effectively taken into 
account\cite{ritz-seckel} by
neglecting the contributions of the light quarks directly
produced in the annihilation process or in 
the hadronization of heavy quarks and gluons,
because these light particles stop inside the medium 
(Sun or Earth) before their decay.
For the case of the Sun,
also the energy loss of the heavy hadrons 
in the solar medium and the energy loss of neutrino
themselves have to be considered\cite{ritz-seckel}.

The differential neutrino flux is then calculated as
\begin{equation}
\frac{dN_\nu}{dE_\nu} =
\frac{\Gamma_A}{4\pi d^2} \sum_{F,f}
B^{(F)}_{\chi f}\frac{dN_{f \nu}}{dE_\nu}  
\end{equation}
where $\Gamma_A$ denotes the annihilation rate,
$d$ is the distance of the detector from the source (i.e. the
center of the Earth or the Sun), $F$ denotes the 
neutralino pair annihilation final states,
$B^{(F)}_{\chi f}$ denotes the branching ratios into
heavy quarks, $\tau$ lepton and gluons in the channel $F$.
The spectra $dN_{f \nu}/dE_{\nu}$ are the differential 
distributions of the neutrinos generated by the $\tau$ and
by hadronization of quarks
and gluons and the subsequent semileptonic decays of the
produced hadrons. A detailed calculation of these spectra
is usually performed by means of a Montecarlo 
simulation\cite{noi_nuflux}.
The spectra due to heavier final states,
i.e. Higgs bosons, gauge bosons and t
quark, can be computed analytically by following their 
decay chain down to the production of a b quark, c quark 
or a tau lepton, where the result of the Montecarlo
simulation can be applied\cite{noi_nuflux,altri_nuflux}.

The neutrino flux may be detected in a neutrino telescope  
by measuring the muons which are produced 
by $\nu_\mu$ and ${\bar \nu}_\mu$ interactions in the 
rock around the detector and then traverse it upwardly. 
Therefore, the signal consists of a flux of up--going
muons, which is computed as
\begin{equation}
\frac{d N_\mu}{d E_\mu}
= N_A \int^\infty_{E_\mu^{\rm th}} d E_\nu  
\int_0^\infty dX \int_{E_\mu}^{E_\nu}
d {E'_\mu }\,\,
P_{\rm surv}(E_\mu,E'_\mu; X)\,\,  
\frac{d \sigma_\nu (E_\nu,E'_\mu)}{d E'_\mu} \,\,
\frac{d N_\nu}{d E_\nu}\, ,
\label{upmu_flux}
\end{equation}
where $X$ is the muon range in the rock,
$d \sigma_\nu (E_\nu,E'_\mu) / d E'_\mu$ is
the charged current cross--section for the
production of a muon of energy $ E'_\mu$ from 
a neutrino of energy $E_\nu$ and
$P_{\rm surv}(E_\mu,E'_\mu; X)$ is the survival
probability that a muon of initial energy $E'_\mu$
will have a final energy $E_\mu$ after propagating
along a path--length $X$ inside the rock which
surrounds the detector. The function
$P_{\rm surv}(E_\mu,E'_\mu; X)$ therefore takes
into account the energy losses of muons in the rock.
In Eq.(\ref{upmu_flux}), $E_\mu^{\rm th}$ is
the detector threshold energy, which for current
neutrino telescopes like MACRO and Baksan is
of about 1-2 GeV, and this is quite suitable for
neutralino indirect detection,
especially for neutralinos lighter than about 100 GeV. 
Large--area neutrino telescopes with higher threshold energies 
(above a few tens of GeV)
are more suitable for heavier relic particles.

Experimentally, one searches for a
statistically significant excess of up--going
muons above the muon flux originated
from atmospheric neutrinos. The different angular
behaviour of the signal with respect to the
atmospheric neutrino background, which has a rather
flat distribution as a function of the zenith angle, 
is the handle for the signal--to--background discrimination. 
Clearly, the flux from the Sun can be pointed at
directly towards the direction of the Sun. In
the case of the flux from the Earth, the process
of accumulation of neutralinos
induces a rather peaked distribution of the
neutrino source around the Earth's center.
Indeed, the angular distribution is 
\begin{equation}
G(\theta) \simeq 4 m_\chi \alpha 
\exp (-2 m_\chi \alpha \sin^2 \theta) 
\end{equation}
where $\theta$ is the zenith angle and
$\alpha = 1.76$ GeV$^{-1}$. This means that
for neutralinos (which are heavier than $\sim$ 25 GeV) the
signal is produced inside a region whose
angular extension is less than about 10 degrees.

The experimental searches at neutrino telescopes
have found no muon excess so far and therefore upper limits on the 
muon flux have been set. The solid line in 
Fig. 3a is the
current most stringent experimental upper limit
from the MACRO Collaboration\cite{MACRO,taup_direct} 
for the neutrino flux from the center of the Earth. Fig. 3b
refers to the flux from the Sun. 

Again, superimposed to the experimental limits,
we show in Fig. 3a (Earth) and Fig. 3b (Sun) 
the susy scatter plot for the
up--going muon signal $\Phi_\mu^{\rm Earth}$
and  $\Phi_\mu^{\rm Sun}$.
The susy configurations have been varied in the ranges 
quoted in Sect. 6.1. The astrophysical parameters are 
$\rho_l = 0.3$ GeV cm$^{-3}$, $v_0 = 220$ Km s$^{-1}$,
$v_{\rm esc} = 550$ Km s$^{-1}$ and the cosmological 
bound $\Omega_\chi h^2 \leq 0.7$ has been applied.

\subsection{Indirect detection: antimatter and gamma rays}

The annihilation process of dark matter neutralinos may take
place also directly in the galactic halo. In this case, many 
different signals other than neutrinos are possible. 
These signals, at variance with the signals previously discussed,
which depend only on local galactic properties,
depend directly on the matter distribution of neutralinos over the whole 
Galaxy. Moreover, the propagation 
inside the Galaxy of the particles which constitute the signal
is perturbed by the Galaxy itself, like, for instance, interactions
with the interstellar medium or, in the case of charged particles,
diffusion in random magnetic fields.

One of the most interesting possibility is the production
of antimatter from neutralino annihilation in the halo. 
The fluxes have been calculated for production of antiprotons,
antideuteron and positrons.

{\em Antiprotons}\cite{noi_pbar,others_pbar} and 
{\em Antideuterons}\cite{dbar} can be produced by
the decay and hadronization of the final state particles
of the annihilation process. The differential rate per unit 
volume and unit time for the production
of $ \bar p$'s from neutralino pair annihilation is defined as
\begin{equation}
q_{\bar p}^{\rm susy}(T_{\bar p}) =
\, <\sigma_{\rm ann} v> 
\left( \frac{ \rho_\chi (\vec r)}{m_\chi} \right)^2\,
 \sum_{F,h}
B^{(F)}_{\chi h} \frac{dN^{h}_{\bar p}}{dT_{\bar p}} \, ,
\label{eq:source}
\end{equation}
where $<\sigma_{\rm ann} v>$ denotes the average over the galactic velocity
distribution function of neutralino pair annihilation cross section
$\sigma_{\rm ann}$ multiplied by the relative velocity $v$ of the annihilating
particles, $T_{\bar p}$ is the antiproton kinetic energy and
$B^{(F)}_{\chi h}$ is the branching ratio for the production of
quarks or gluons $h$ due to the decay of the particles produced by
neutralino annihilation into the final state $F$.
Finally, $dN^h_{\bar p}/dT_{\bar p}$ is the differential energy distribution
of the antiprotons generated by hadronization of quarks and gluons.
Notice that the rate depends on the square of the
mass distribution function of neutralinos in the galactic halo
$\rho_\chi (\vec r)$. The rate of production of $\bar D$ is clearly
analogous to Eq.(\ref{eq:source}).

After being produced, antimatter propagates inside the Galaxy
and it experiences both diffusion in the galactic magnetic field
and energy losses, due to ionization, scattering, collision 
and others. The propagation of antiprotons in the galactic
medium is properly calculated in a diffusion model where
the Galaxy is described as composed  of two zones, one for
the disk and the other for the halo. The diffusion
equation which governs the behaviour of antiprotons is
\begin{equation}
\vec{\nabla} \cdot \left( K \; \vec{\nabla} \, \mbox{$\psi_{\bar{\rm
p}}$}\right) \; - \;
2 h \delta(z) \, \Gamma_{\bar{\rm p}} \, \mbox{$\psi_{\bar{\rm p}}$}\; + \;
2 h \delta(z) \, \mbox{$q_{\bar{\rm p}}^{\rm susy}$}(r) \; - \;
2 h \delta(z) \, \frac{\partial}{\partial E} \left\{ b(E)
\mbox{$\psi_{\bar{\rm p}}$}\right\}
\; = \; 0 \;\; ,
\label{DIFFUSION_PBAR_DISK}
\end{equation}
where $\psi_{\bar p}$ denotes the antiproton density,
$K$ is the diffusion coefficient, $h$ is the height
of the galactic disk and $\Gamma_{\bar{\rm p}}$
is the collision probability with the
interstellar medium and $b(E)$ describes 
energy losses.
An analogous equation holds for the antideuterons.

Solution of the diffusion equations gives the antiproton
(or antideuterons) flux at the heliosphere boundaries
(interstellar flux) as
\begin{equation}
\Phi_{\bar{\rm p}} ( \odot , T_{\bar{\rm p}} ) \,\, = \,\,
<\sigma_{\rm ann} v> \,
\frac{v_{\bar p}}{4 \pi} \,  
\left ( \frac{ \rho_l}{m_\chi} \right )^2 \,
\psi_{\bar{\rm p}}^{\rm eff} ( \odot , T_{\bar{\rm p}} )
\cdot \sum_{F,h} B^{(F)}_{\chi h} \frac{dN^{h}_{\bar p}}{dT_{\bar p}}\, ,
\label{SUSY_ANALYSIS}
\end{equation}
where $\psi_{\bar{\rm p}}^{\rm eff} ( \odot , T_{\bar{\rm p}} ) $ 
is obtained by solving Eq.(\ref{DIFFUSION_PBAR_DISK}). 

Antimatter subsequently enters the heliosphere where it
propagates against the solar wind before arriving at the Earth.
The effect induced by the solar wind (solar modulation) is
quite important at low kinetic energies, and introduces a time
dependence into the calculation, since it is correlated
with the 11 year solar cycle.

Both antiprotons and antideuterons can be
produced also by the interaction of primary cosmic rays on
the interstellar medium. This secondary antimatter fluxes
constitute a background for the susy signal. The background
fluxes have a different energy behaviour as compared to the
the ones of susy origin. In particular, the low energy tail
of the energy spectrum is the most interesting place to
look at, since the signals have a somewhat flat behaviour, 
while the secondary fluxes are depressed by 
kinematical reasons \cite{secondary}. This effect is stronger for
antideuterons than for antiprotons. Therefore, 
the antideuteron flux presents some advantages 
of discrimination over background with respect to 
the antiproton flux, even if its smallness makes 
the detection harder to be achieved. 

Fig. 4 show the antiproton flux\cite{noi_pbar} $\Phi_{\bar p}$
vs. the neutralino mass for the susy configurations
introduced in Sect. 6.1. The flux is calculated for 
a $\bar p$ kinetic energy $T_{\bar p} = 0.24$ GeV, to
conform to the first BESS energy bin, and for a solar
modulation phase close to the BESS data--taking periods.
The horizontal
line is the present upper limit derived from the
BESS 95 and BESS 97 data\cite{BESS97}. We notice that antiproton
measurement are already able to exclude some
susy configurations which would imply a too large
$\bar p$ flux at low kinetic energies.

Fig. 5 shows the antideuteron flux\cite{dbar} $\Phi_{\bar D}$
vs. the neutralino mass for the same susy models.
The flux is calculated for a $\bar D$ kinetic energy 
$T_{\bar D} = 0.24$ GeV/nucleon, at the maximum
of solar activity. The horizontal line
is the sensitivity level which could be achieved by
the AMS detector during the flight on space station.

{\em Positrons}\cite{suede_positron} are produced again from the decay chain of
the neutralino annihilation products. It is also possible
to produce directly a pair of monocromatic electron--positron
pair. The branching ratio for this process is usually very
small, but some susy models can present a production rate
strong enough to be at the level of the detector 
sensitivity\cite{suede_positron}. 
The calculation of the positron flux is analogous to
the one for antiprotons or antideuterons, with the inclusion 
of additional energy loss mechanisms:
inverse Compton scattering on starlight and cosmic microwave
background, and synchrotron radiation emission in the galactic
magnetic field. The IS positron flux is then affected by the 
solar wind before coming to the Earth where it can be detected.
Fig. 6, taken from Ref.\cite{suede_positron},
shows a scatter plot of the positron flux $\Phi_{e+}$
integrated in the energy range 8.9 -- 14.8 GeV, to
conform to one of the HEAT data bins.
The susy parameters have been varied in the ranges:
$0\;\mbox{GeV} \leq |M_2| \leq  5000\;\mbox{GeV},\; 
0\;\mbox{GeV} \leq |\mu| \leq  5000\;\mbox{GeV},\;
0\;\mbox{GeV} \leq m_A \leq  10000\;\mbox{GeV},\; 
100\;\mbox{GeV} \leq m_0 \leq  30000\;\mbox{GeV},\;
-3 \leq {\rm A} \leq +3,\; 1 \leq \tan \beta \leq 60$. 
Only configuration with $0.025 \leq \Omega_\chi h^2 \leq 1$ are
shown on the plot. The horizontal dashed line denotes the value
of the background positrons of secondary origin calculated
in Ref.\cite{moskalenko}. The horizontal band represents the HEAT 94
data\cite{HEAT}.

The last possibility we consider here is the production of 
{\em diffuse gamma rays} or a {\em gamma line} 
(see Refs.\cite{gamma,gamma_bbm,line_bba,bub,suede_clump,gamma_annihil} 
and references quoted therein).
Diffuse photons are produced mainly through the decay of neutral
pions originated from the hadronization of the
neutralino annihilation products. A monochromatic gamma line, instead,
is produced through the loop--processes 
$\chi \chi \longrightarrow \gamma \gamma$ and 
$\chi \chi \longrightarrow Z \gamma$.
In this case, the gamma line would constitute a particularly
nice signature, since it is practically background free.

In both cases, the fluxes are usually rather low and,
in order to have fluxes at the level of the detector 
sensitivities, some matter over-density is needed, like
for example a singular dark matter halo\cite{gamma_bbm,bub} 
or a clumpy matter distribution\cite{suede_clump}. 
Fig. 7, taken from Ref.\cite{suede_clump},
shows a scatter plot of the gamma ray flux  
$\Phi_{cont, \gamma}$ integrated above $E_\gamma = 1$ GeV. 
The susy parameters have been varied in the same
ranges defined above for the positron flux.
Only configurations with $0.025 \leq \Omega_\chi h^2 \leq 0.5$ are
shown on the plot. The horizontal line denotes the integrated
gamma ray flux measured by EGRET\cite{EGRET}.
Fig. 8 from Ref. \cite{bub} shows the perspectives of a number of Air
Cherenkov Telescopes for the measurements of the expected $\gamma$--ray
line.

\section{Conclusions}
As we have seen above, the question of dark matter in the Universe presents a
large number of intriguing facets of relevance for cosmology, astrophysics and
particle physics. It represents a field  in a very fast expansion, because
of an impressive development in experimental activities as well as in
theoretical investigations. Let us mention just some of the most promising
avenues: a) Further observations and analyses of high--$z$ Supernovae, of the 
Cosmic Microwave Background Radiation, and of the time evolution of the number
density of clusters are expected to provide more conclusive information on 
$\Omega_m$ and $\Omega_{\Lambda}$. b) New numerical simulations of cosmological
structures 
should give a unique information about the (hot/cold) composition of dark 
matter
and about crucial details on the dark matter density profile in single
galaxies. c) Further accumulation of data in WIMP direct detection are
expected to play a fundamental role in the process of the 
identification of dark matter
particle candidates. Present experimental results have already allowed the 
pinning down of a sector of the supersymmetric parameter space, part of which
can be explored at accelerator and by WIMP indirect searches. 
It has been proved that the relevant relic
neutralinos might behave as major components of cold dark matter. 
No doubt that the connection of particle physics with the dark matter problem 
in the Universe is one of the most exciting and far--reaching field in 
astroparticle physics.

\bigskip
\bigskip
\begin{center}
{\bf Acknowledgments}
\end{center}
This work was supported by DGICYT under grant number 
PB95--1077 and by the TMR network grant ERBFMRXCT960090 of 
the European Union.

\newpage
\begin{figure}[t]
\hbox{
\psfig{figure=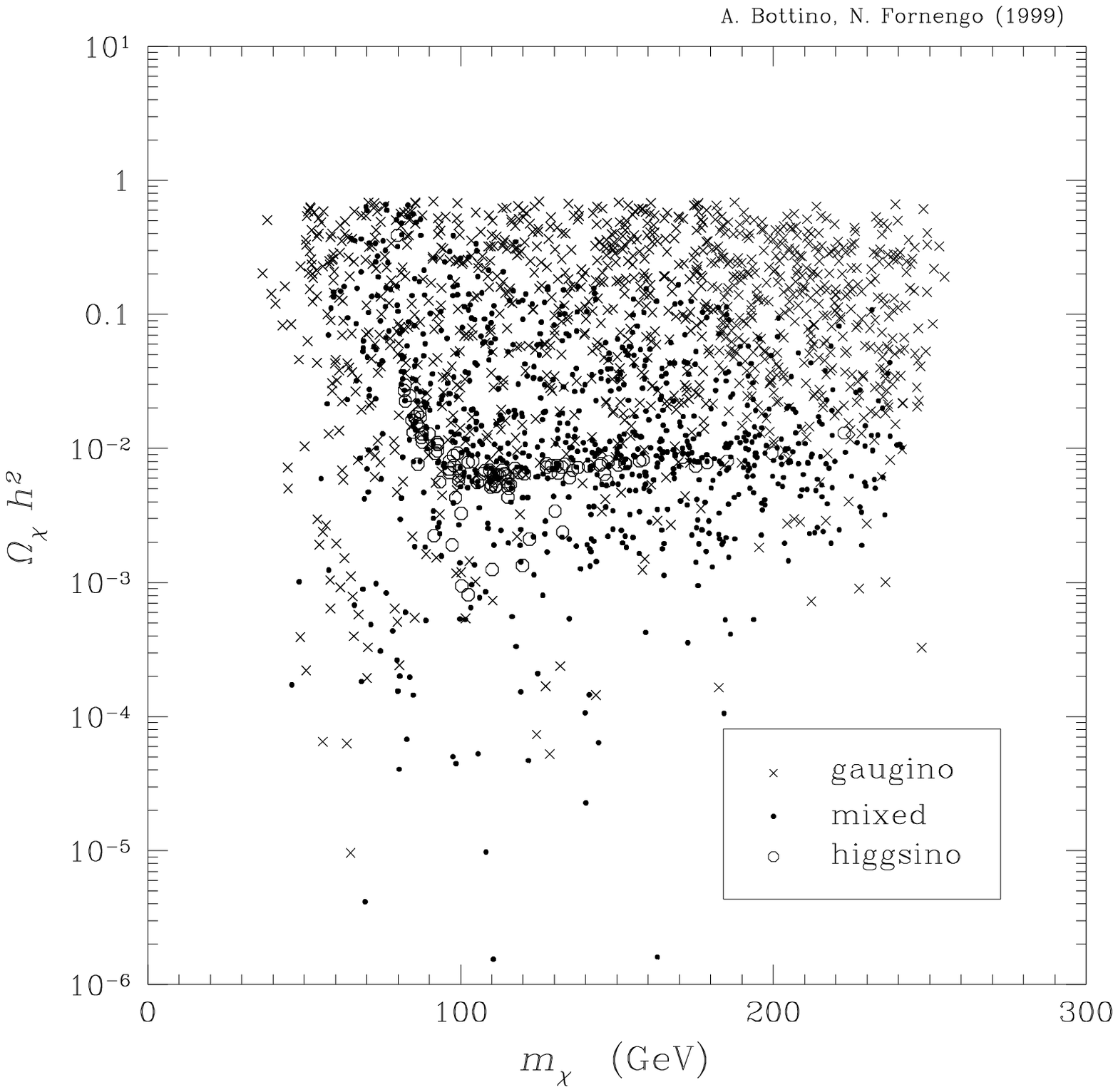,width=8.2in,bbllx=40bp,bblly=160bp,bburx=700bp,bbury=660bp,clip=}
}
{FIG. 1. Neutralino relic abundance $\Omega_\chi h^2$ as a function of
the neutralino mass $m_\chi$, evaluated in the MSSM. 
Only configurations which provide a relic abundance
not in contrast with the age of the Universe 
(i.e. $\Omega_\chi h^2 < 0.7$) are shown.
}
\end{figure}

\newpage
\begin{figure}[t]
\hbox{
\psfig{figure=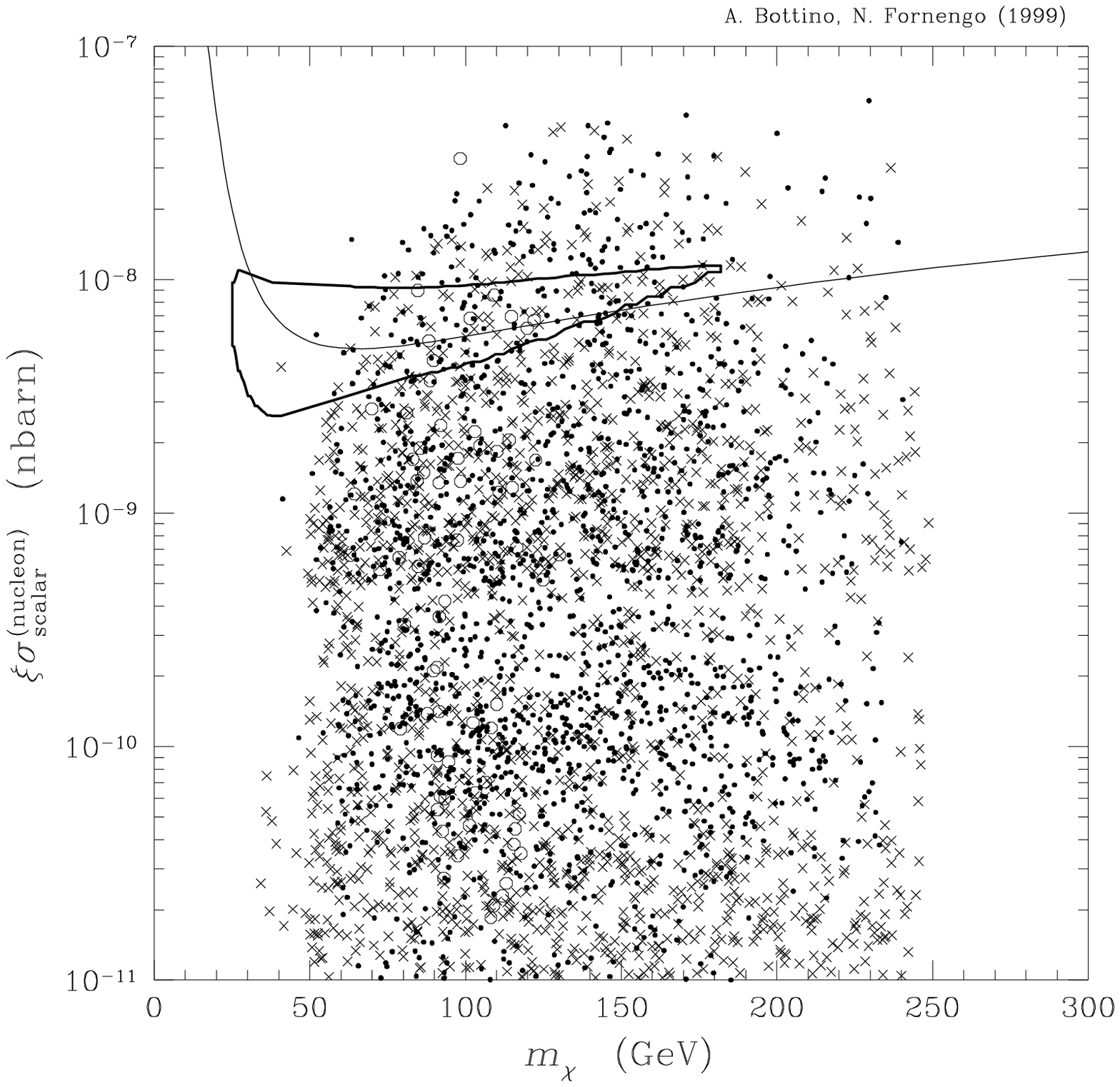,width=8.2in,bbllx=40bp,bblly=160bp,bburx=700bp,bbury=660bp,clip=}
}
{FIG. 2. Neutralino--nucleon cross section 
$\sigma^{(\rm nucleon)}_{\rm scalar}$
times the fractional amount of neutralino dark matter $\xi$, as
a function of the neutralino mass $m_\chi$.
The figure refers to the value $\rho_l = 0.3$ GeV cm$^{-3}$ for the total
local dark matter density. The solid line is the present upper limit
at 90\% C.L.\cite{DAMA_uplim}. The closed contour delimits the region compatible 
with the annual modulation effect\cite{DAMA,bellietal}. The scatter plot shows the
quantity $\xi \sigma^{(\rm nucleon)}_{\rm scalar}$ evaluated in the
MSSM. Different neutralino compositions are shown with different symbols:
crosses for gauginos, open circles for higgsinos and dots for mixed
neutralinos.
}
\end{figure}

\newpage
\begin{figure}[t]
\hbox{
\psfig{figure=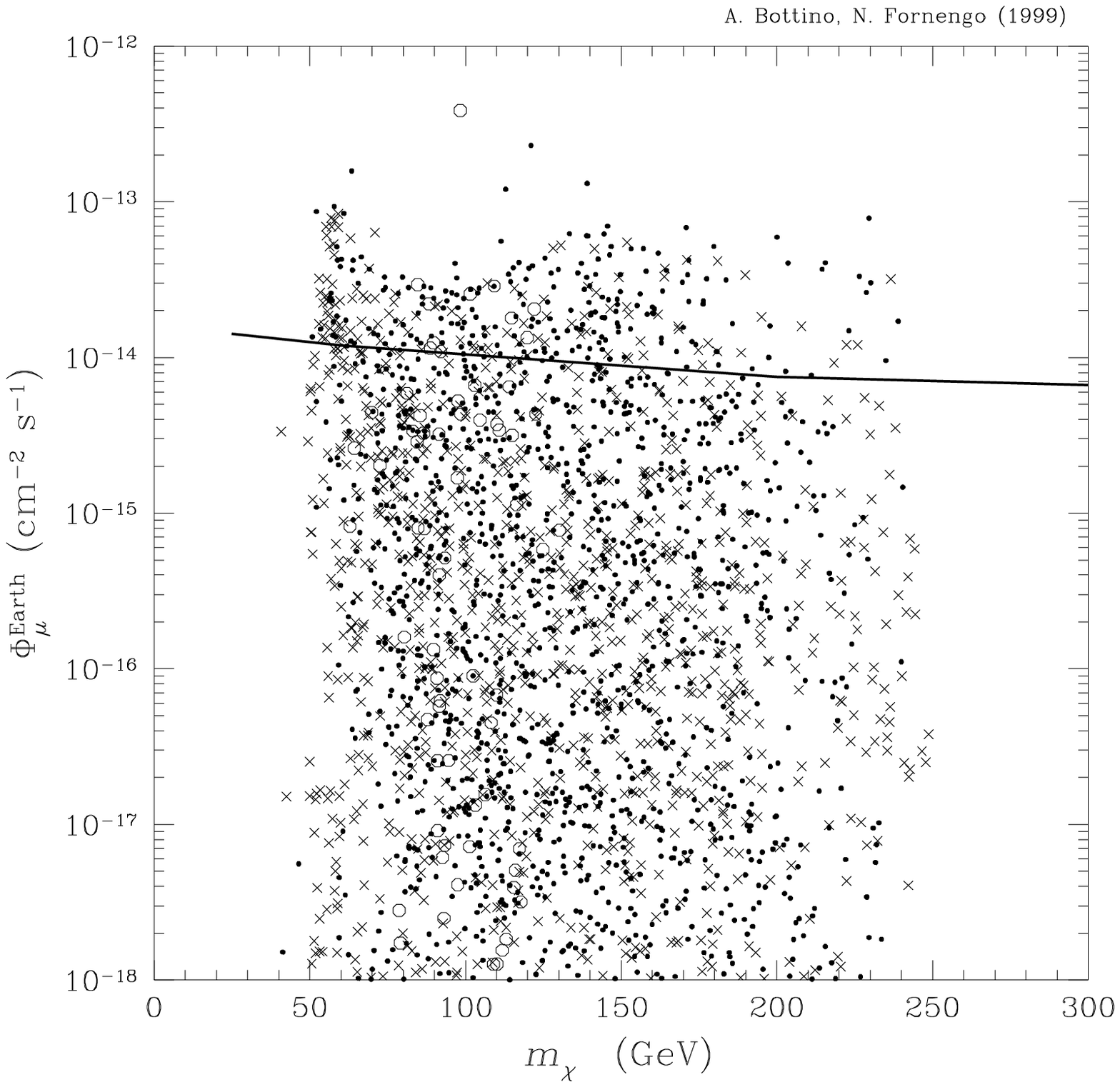,width=8.2in,bbllx=40bp,bblly=160bp,bburx=700bp,bbury=660bp,clip=}
}
{FIG. 3a. Flux of up--going muons $\Phi_\mu^{\rm Earth}$ from
neutralino annihilation in the Earth, plotted as a function of
$m_\chi$. The solid line denotes the present upper limit\cite{MACRO}.
Different neutralino compositions are shown with different symbols:
crosses for gauginos, open circles for higgsinos and dots for mixed
neutralinos.
}
\end{figure}

\newpage
\begin{figure}[t]
\hbox{
\psfig{figure=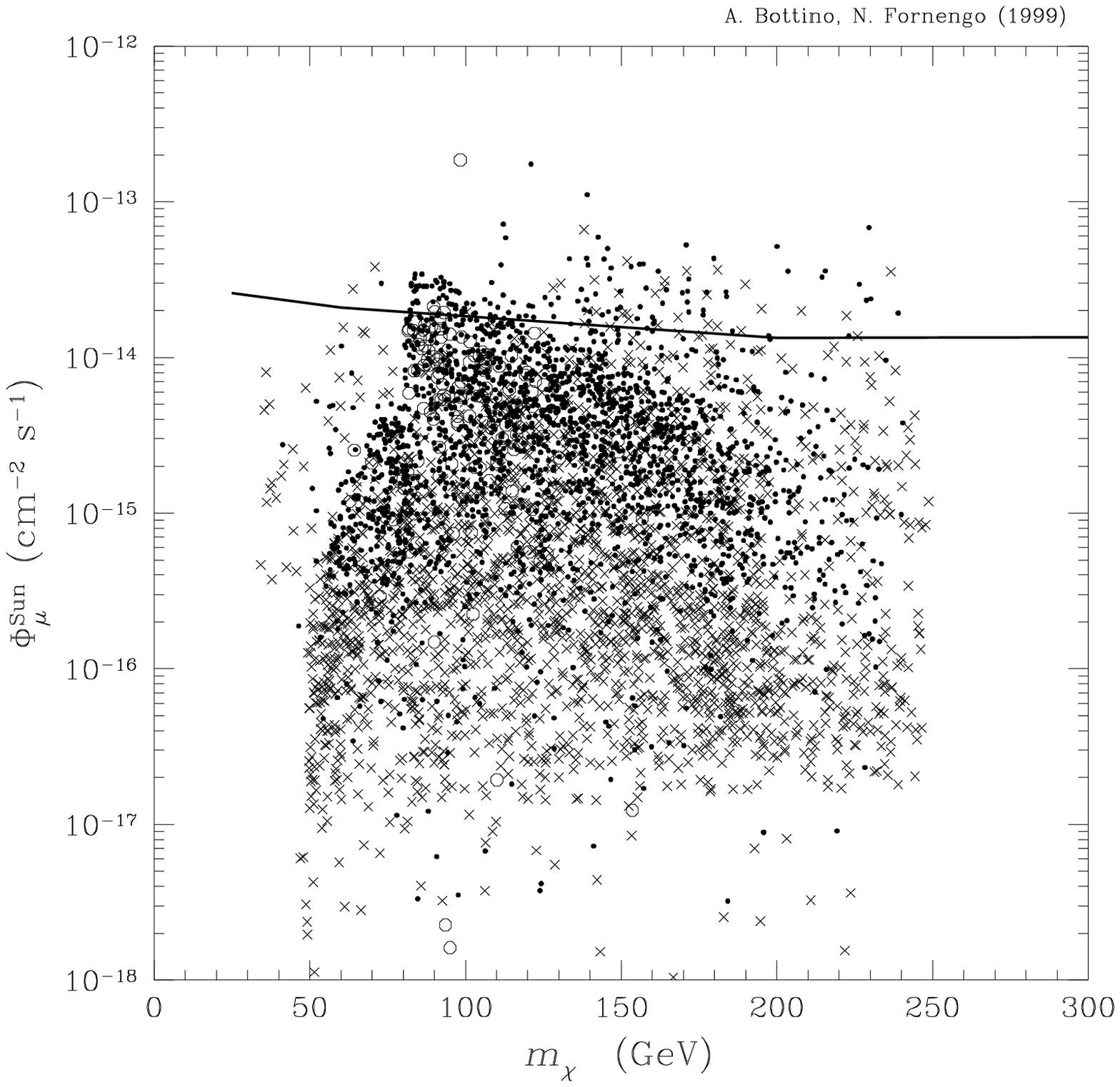,width=8.2in,bbllx=40bp,bblly=160bp,bburx=700bp,bbury=660bp,clip=}
}
{FIG. 3b. Flux of up--going muons $\Phi_\mu^{\rm Sun}$ from
neutralino annihilation in the Sun, plotted as a function of
$m_\chi$. The solid line denotes the present upper limit\cite{MACRO}.
Different neutralino compositions are shown with different symbols:
crosses for gauginos, open circles for higgsinos and dots for mixed
neutralinos.
}
\end{figure}

\newpage
\begin{figure}[t]
\hbox{
\psfig{figure=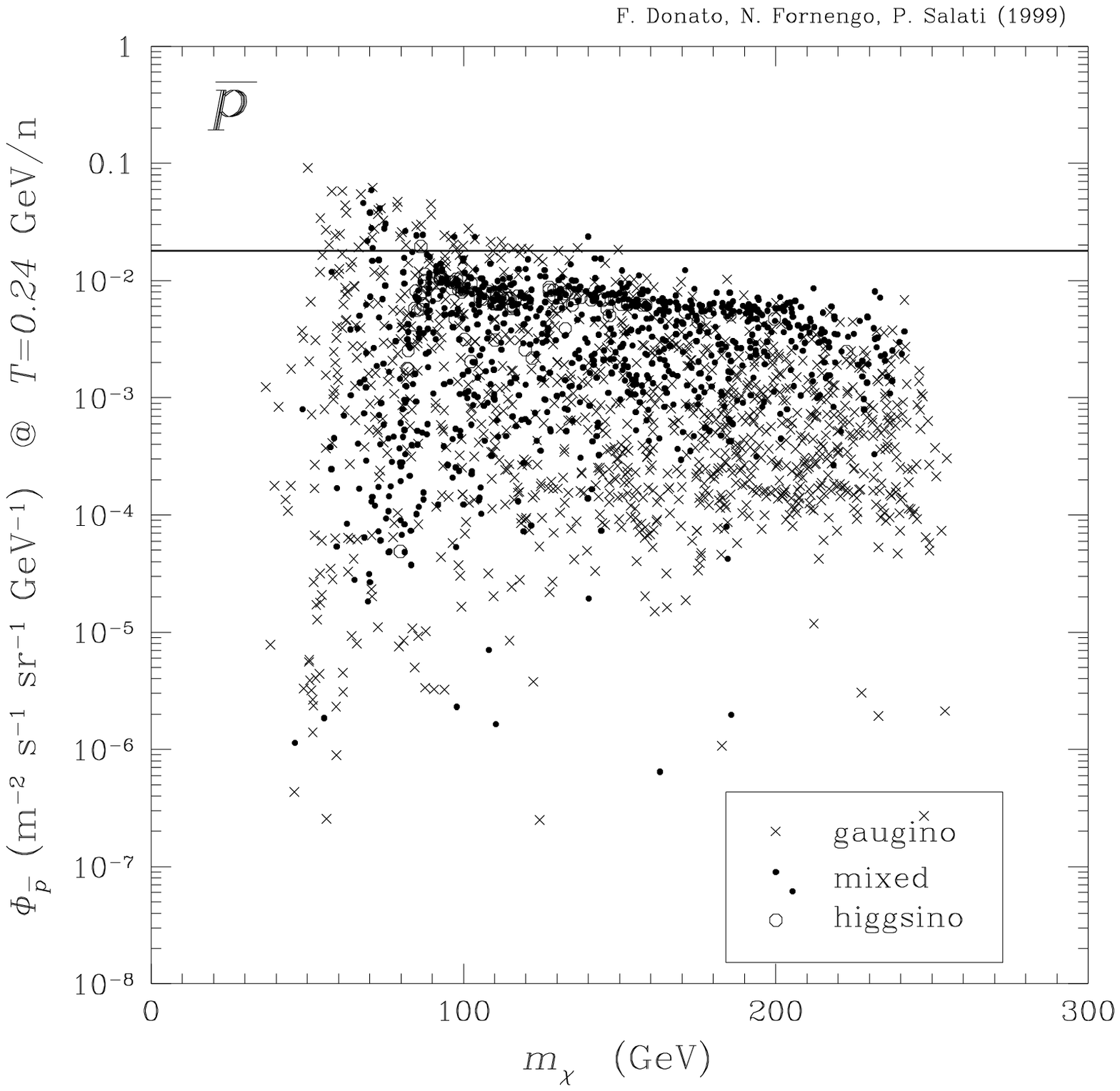,width=8.2in,bbllx=40bp,bblly=160bp,bburx=700bp,bbury=660bp,clip=}
}
{FIG. 4. Antiproton flux $\Phi_{\bar p}$
as a function of $m_\chi$ for 
a $\bar p$ kinetic energy $T_{\bar p} = 0.24$ GeV
and for the solar minimum. The horizontal
line is the present upper limit derived from the
BESS 95 and BESS 97 data\cite{BESS97}.
}
\end{figure}

\newpage
\begin{figure}[t]
\hbox{
\psfig{figure=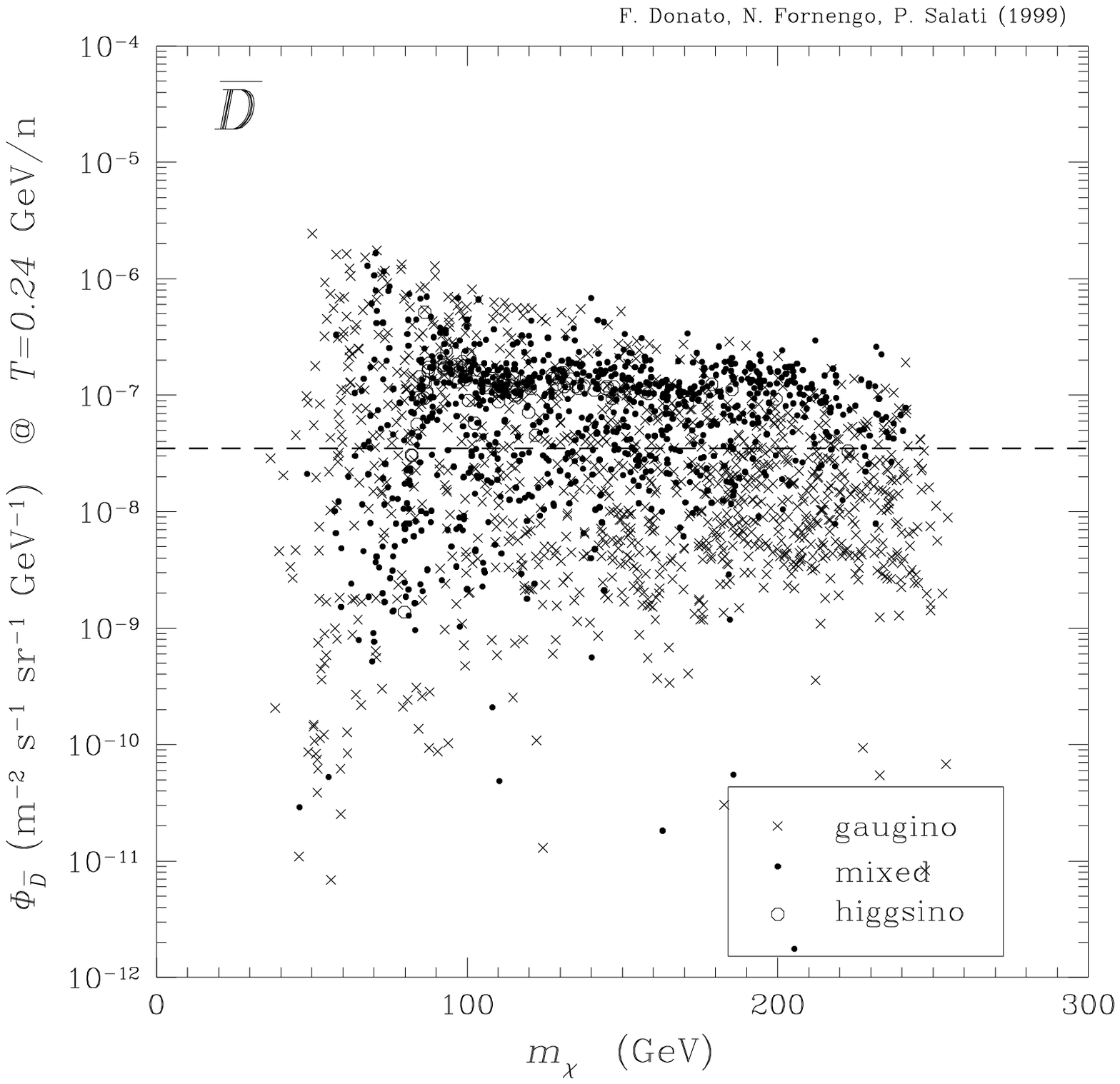,width=8.2in,bbllx=40bp,bblly=160bp,bburx=700bp,bbury=660bp,clip=}
}
{FIG. 5.  Antideuteron flux $\Phi_{\bar p}$
as a function of $m_\chi$ for 
a $\bar D$ kinetic energy $T_{\bar D} = 0.24$ GeV
and for the solar maximum. The horizontal
line is the sensitivity that could be achieved by
AMS during the flight on the space station \cite{dbar}.
}
\end{figure}

\newpage
\begin{figure}[t]
\hbox{
\psfig{figure=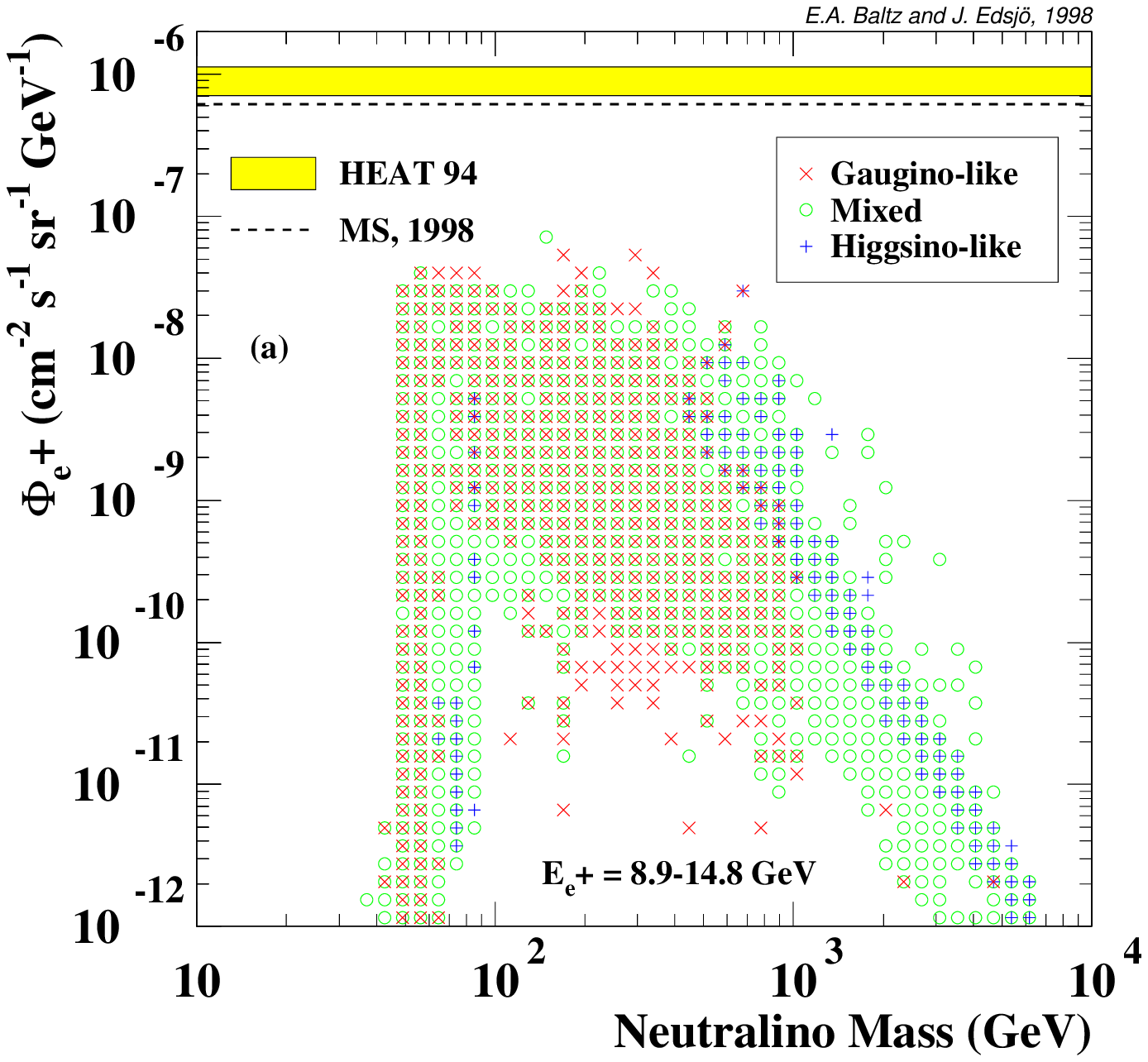,width=8.2in,bbllx=0bp,bblly=0bp,bburx=600bp,bbury=400bp,clip=}
}
{FIG. 6. Positron flux $\Phi_{e+}$
integrated in the energy range 8.9 -- 14.8 GeV vs. $m_\chi$, for
configuration with $0.025 \leq \Omega_\chi h^2 \leq 1$.
 The horizontal dashed line denotes the value
of the background positrons of secondary origin\cite{moskalenko}. 
The horizontal band represents the HEAT 94 data.
}
\end{figure}

\newpage

\begin{figure}[t]
\hbox{
\psfig{figure=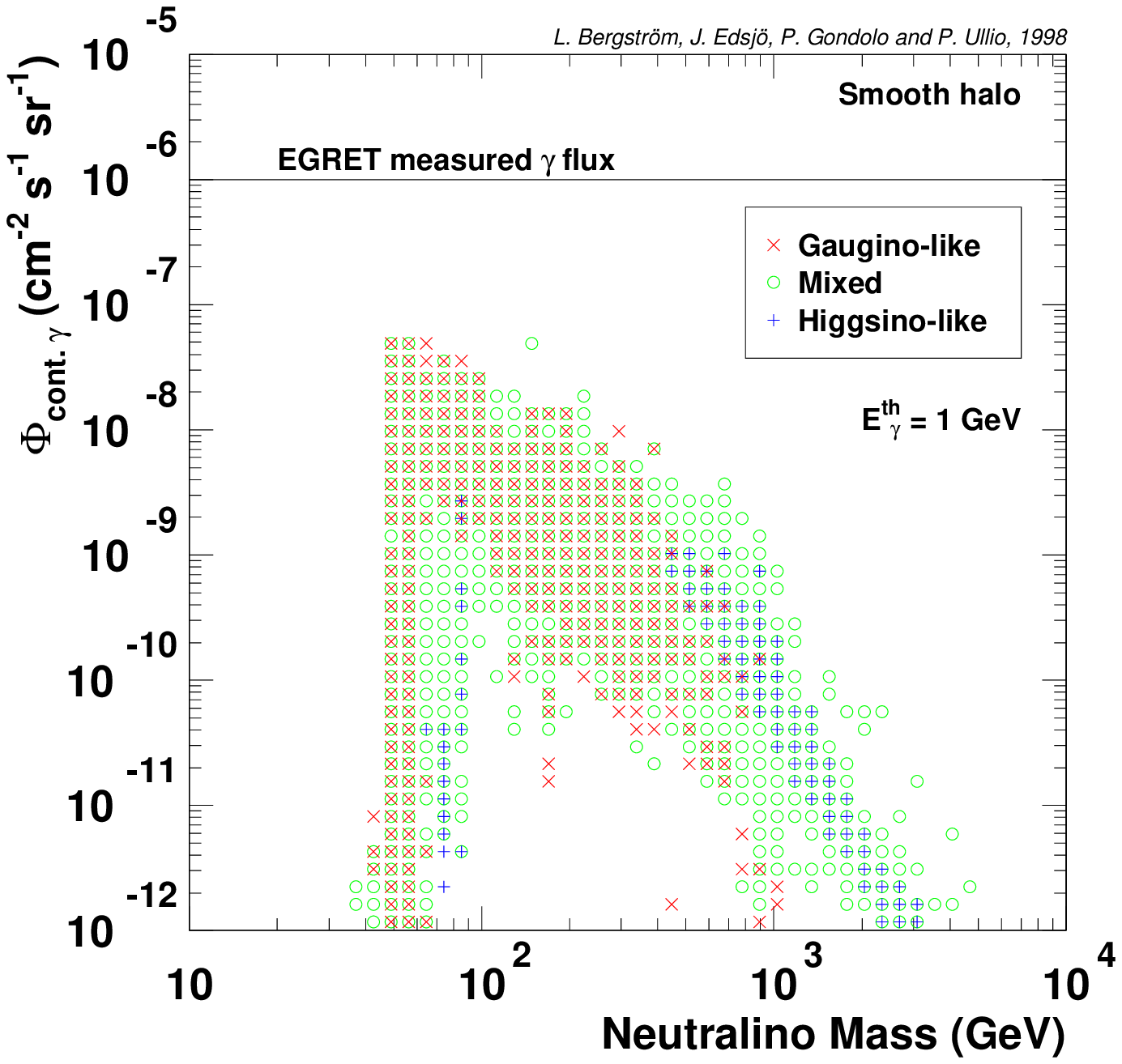,width=8.2in,bbllx=0bp,bblly=0bp,bburx=600bp,bbury=400bp,clip=}
}
{FIG. 7. Gamma ray flux  
$\Phi_{cont, \gamma}$ integrated above $E_\gamma = 1$ GeV
vs. $m_\chi$, for
configurations with $0.025 \leq \Omega_\chi h^2 \leq 0.5$.
The horizontal line denotes the integrated
gamma ray flux measured by EGRET\cite{EGRET}.
}
\end{figure}

\newpage

\begin{figure}[t]
\hbox{
\psfig{figure=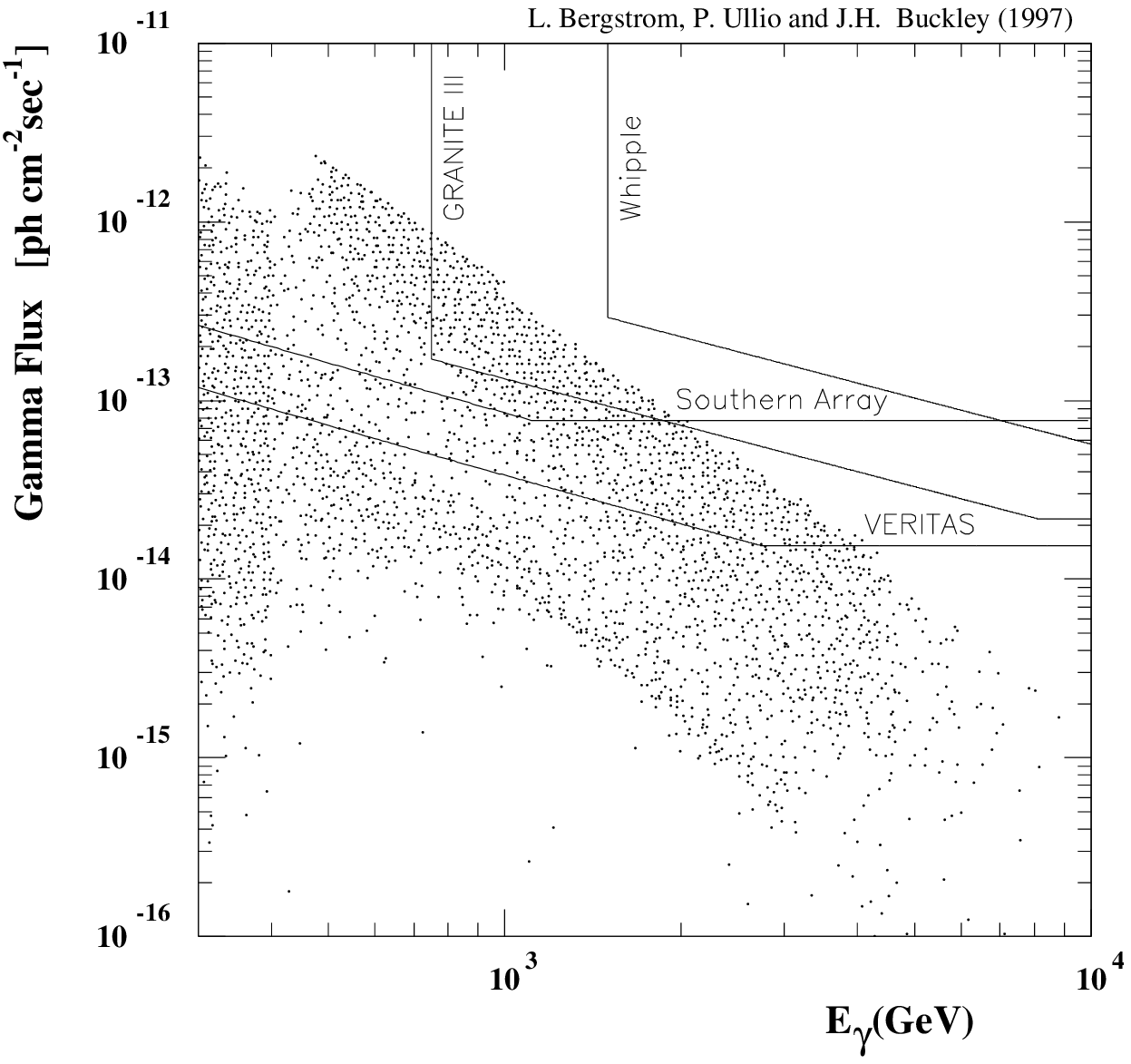,width=8.2in,bbllx=0bp,bblly=0bp,bburx=600bp,bbury=400bp,clip=}
}
{FIG. 8. Gamma ray flux inside a $10^{-5}$ sr angular cone which
contains the galactic center, for the density matter distribution
of Ref.\cite{nfw1}. The flux refers to the contributions
of the $2\gamma$ and $\gamma Z$ annihilation lines for heavy
neutralinos. The solid lines denote the estimated sensitivities 
for different atmospheric Cherenkov detectors.
}
\end{figure}


\bigskip
\bigskip


\begin{thebibliography}{99}
\bibitem{weinberg} S. Weinberg, {\it Gravitation and Cosmology} (J. Wiley 
\& Sons, 1972).

\bibitem{kt} E.W. Kolb and M.S. Turner, {\it The Early Universe} 
(Addison--Wesley,1988).

\bibitem{peebles} P. J. E. Peebles, {\it Principles of Physical Cosmology}, 
(Princeton University Press, 1993).

\bibitem{roos} M. Roos, {\it Introduction to Cosmology} (J. Wiley \& Sons, 
1994).

\bibitem{kolb} E.W. Kolb, Proceedings of the Fourth School on 
{\it Non--accelerator Particle Astrophysics} 
(Editors E. Bellotti, R.A. Carrigan Jr., 
G. Giacomelli and N. Paver, World Scientific, 1996).

\bibitem{freedman} W.L. Freedman, talk given at CAPP 98, CERN, Geneva, 
June 1998, {\sl http://wwwth.cern.ch /capp98/programme.html}.

\bibitem{cdkk} B. Chaboyer, P. Demarque, P.J. Kernan and L.M. Krauss, 
Ap.J. 494 (1998) 96.

\bibitem{madore} B.F. Madore et al., Nature 395 (1998) 47.

\bibitem{sap} M. Spiro, E. Aubourg and 
Palanque--Delabrouille, Nucl. Phys. B (Proc. Supplements) 70 (1999) 14. 

\bibitem{wf} S.D.M. White, J.F. Navarro, A. Evrard and C.S. Frenk, Nature 366 
(1993) 429; D. White and A. Fabian, Mon. Not. Roy. Astron. Soc. 272 
(1995) 72; L. Lubin, R. Cen, N.A. Bahcall and J.P. Ostriker, 
Ap.J. 460 (1996) 10.

\bibitem{nucl} K. Olive, Nucl. Phys. B (Proc. Supplements) 70 (1999) 521.

\bibitem{peebles2} P.J.E. Peebles, Proceedings of the
      Smithsonian debate {\it Is Cosmology Solved?},
      (PASP, to appear), {\sl astro--ph/9810497}.

\bibitem{perl1} S. Perlmutter, Nature 391 (1998) 51.

\bibitem{perl2} S. Perlmutter, talk given at CAPP 98, CERN, Geneva, June 1998, 
{\sl http://wwwth.cern.ch /capp98/programme.html}.

\bibitem{reiss} A.G. Riess et al., {\sl astro--ph/9805201} (to appear in
Astron. J.).

\bibitem{new} For a recent review see, for instance, C.R. Lawrence, D. Scott
0and M. White, report UBC--COS--98--04, {\sl astro--ph/9810446}, 
and references therein. 

\bibitem{lineweaver} C.H. Lineweaver, {\sl astro--ph/9805326}
(to appear in Astroph. J. Lett.).

\bibitem{neta} N.A. Bahcall and X. Fan, {\sl astro--ph/9804082} 
(to appear in National Academy of Sciences Proc.).

\bibitem{pac} B. Pacz\'ynski, Ap. J. 304 (1986) 1. 

\bibitem{spiro} M. Spiro, review talk at the 19th Texas Symposium, 
Paris, December 1998.

\bibitem{stubbs} C. Stubbs, talk given at the 19th Texas Symposium, 
Paris, December 1998. 

\bibitem{eros} R. Ansari et al. (EROS Collaboration), Astronomy and
Astrophysics 314 (1996) 94. 

\bibitem{macho} C. Alcock et al. (MACHO Collaboration), Ap. J. 299 (1996) 84. 

\bibitem{FFG} K. Freese, B. Fields and D. Graff, {\sl astro-ph/9901178}.

\bibitem{jetzer} F. De Paolis, G. Ingrosso, Ph. Jetzer and
M. Roncadelli, {\sl astro-ph/9901033}.

\bibitem{solar} A. Yu. Smirnov, talk at Neutrino 98, Takayama, June 1998,
{\sl hep--ph/9809481}.

\bibitem{smy} M.B. Smy, hep-ph/9903034.

\bibitem{sk} Y. Fukuda et al. (Super--Kamiokande Collaboration), 
     Phys. Rev. Lett. 81 (1998) 1562. 

\bibitem{macro} F. Ronga (MACRO Collaboration), talk at Neutrino 98, 
Takayama, June 1998, {\sl hep--ph/9810008}.

\bibitem{habig} A. Habig, hep-ph/9903047.

\bibitem{lsnd} C. Athanassopoulos et al., Phys. Rev. C54 (1996) 2685; 
Phys. Rev. Lett. 81 (1998) 81. 

\bibitem{prim} J.R. Primack and M.A.K. Gross,
Proceedings of the X$^{\mbox{\rm th}}$ Rencontres de Blois, 
{\it The Birth of Galaxies}, June 1998, {\sl astro--ph/9810204}.

\bibitem{peccei} R.D. Peccei and H.R. Quinn, Phys. Rev. D16 (1977) 1791.

\bibitem{axion} G. Raffelt, talk at Neutrino 98, Takayama, 1998,
{\sl hep--ph/9806506}.

\bibitem{neutralino} Because of the relevance of and interest in the
	subject of neutralino dark matter, a complete list of references
	would be too long to be fitted in this short review. We will
        therefore refer here to some of the more recent relevant papers,
	where references to older papers can be found.
        An extensive compilation of references previous to 1996 can also
        be found in G. Jungman, M. Kamionkowski and K. Griest, 
        Phys. Rep. 267 (1996) 195.  


\bibitem{snu} T. Falk. K.A. Olive and M. Srednicki,
	        Phys. Lett. B339 (1994) 248;	
              L.J. Hall, T. Moroi and H. Murayama,
                Phys. Lett. B424 (1998) 305.

\bibitem{gravitino} E.J. Chun, H.B. Kim and J.E. Kim, 
                      Phys. Rev. Lett. 72 (1994) 1956;
		    A. Borgani, A. Masiero and M. Yamaguchi,
		      Phys. Lett. B386 (1996) 189.	

\bibitem{messenger} S. Dimopoulos, G.F. Giudice and A. Pomarol,
		      Phys. Lett. B389 (1996) 37;
		    T. Han and R. Hempfling, 
		      Phys. Lett. B415 (1997) 161.

\bibitem{axino} S.A. Bonometto, F. Gabbiani and A. Masiero,
		  Phys. Rev. D 49 (1994) 3918.  


\bibitem{qballs} A. Kusenko and M. Shaposhnikov,
		   Phys.Lett. B418 (1998) 46;
		 A. Kusenko, V. Kuzmin, M. Shaposhnikov and P.G. Tinyakov,
		   Phys. Rev. Lett. 80 (1998) 3185.

\bibitem{zillas} D.J.H. Chung, E.W. Kolb and A. Riotto,
		  Phys. Rev. Lett. 81 (1998) 4048.


\bibitem{susy} H.P. Nilles, Phys. Rep. 110 (1984) 1;
H.E. Haber and G.L. Kane, Phys. Rep. 117 (1985) 75; 
R. Barbieri Riv. Nuovo Cim. 11 (1988) 1.


\bibitem{sugra}
L. E. Iba{\~n}ez and C. Lopez, Nucl. Phys.  B233 (1984) 511;
J. Ellis and F. Zwirner, Nucl. Phys. B338 (1990) 317; 
M. Drees and M. M. Nojiri, Nucl. Phys. B369 (1992) 54;
R. Arnowitt and P. Nath, Lectures at VII J. A. Swieca 
Summer School, Campos do Jordao, Brazil, 1993, {\sl hep-ph/9309277}; 
J. L. Lopez, D.V. Nanopoulos and A. Zichichi, Riv. Nuovo Cim. (1994) 1; 
W. de Boer, Prog. Part. Nucl. Phys. (1994) 201; 
G. L. Kane, C. Kolda, L. Roszkowski and J. D. Wells, Phys. Rev.
D 49 (1994) 6173; 
R. Rattazzi and U. Sarid, Phys.Rev D 53 (1996) 1553;
X. Wang, J. Lopez and D. Nanopoulos,
Phys. Lett. B348 (1995) 105;
V. Barger, M. S. Berger and P. Ohmann, Phys. Rev. D 49 (1994) 4908. 


\bibitem{sugra_dm} V. Berezinsky, A. Bottino, J. Ellis, N. Fornengo,
G. Mignola and S. Scopel, Astropart. Phys. 5 (1996) 1; 
Astropart. Phys. 5 (1996) 333; R. Arnowitt and P. Nath, 
Phys. Rev. D56 (1997) 2820.

\bibitem{noi_omega}  
A.Bottino, V.de Alfaro, N.Fornengo, G.Mignola and M. Pignone, 
Astropart. Phys. 2 (1994) 67.

\bibitem{omegah2}  
R. Arnowitt and P. Nath, Phys. Lett B299 (1993); H. Baer
and M. Bhrlik, Phys. Rev. D 53 (1996) 597.


\bibitem{coann} K. Griest and D. Seckel, Phys. Rev. D 43 (1991) 3191;
S. Mizuta and M. Yamaguchi, Phys. Lett. B 298 (1993) 120;
M. Drees and M. Nojiri, Phys. Rev. D 47 (1993) 376;
J. Ellis, T. Falk and K. A. Olive, Phys. Lett. B444 (1998) 367.

\bibitem{suede_coann} J. Edsj\"o and P. Gondolo, Phys. Rev. D 56 (1997) 1879. 

\bibitem{LEP}  V. Ruhlmann--Kleider (DELPHI Collaboration),
                    presentation at the LEPC Conference, November 1998;
                    R. Clare (L3 Collaboration), {\em ibid}. 


\bibitem{bsg_exp} S. Glenn (CLEO Collaboration), preprint CLEO CONF 98--17, 1998,
                International Conference on High Energy Physics, 
                Vancouver, 1998, paper 1011;
                R. Barate et al. (ALEPH Collaboration), CERN preprint
                CERN--EP/98--044, 1998.


\bibitem{bsg_theo}  S. Bertolini, F. Borzumati, A. Masiero and G. Ridolfi, 
                    Nucl. Phys. B353 (1991) 591;
                     R. Barbieri and G.F. Giudice, Phys. Lett. B309
                    (1993) 86;
                     R. Garisto and J.N. Ng, Phys. Lett. B315 (1993) 372;
 F.M. Borzumati, M. Drees and M.M. Nojiri, 
                    Phys. Rev. D 51 (1995) 341;
 J. Wu, R. Arnowitt and P. Nath, Phys. Rev.  D 51 (1995) 1371;
 V. Barger, M.S. Berger, P. Ohmann and R.J.N. Phillips, Phys.
                Rev. D 51 (1995) 2438;
 K. Chetyrkin, M. Misiak and M. M\"unz, Phys. Lett. 
                    B400 (1997) 206;
 M. Ciuchini, G. Degrassi, P. Gambino and G.F. Giudice,
                    CERN preprint CERN--TH/97--279, 1997,
                    {\sl hep-ph/9710335};
 A. Czarnecki and W.J. Marciano, 
              Brookhaven National Laboratory
              preprint BNL--HET--98/11, 1998, {\sl hep-ph/9804252};
 M. Ciuchini, G. Degrassi, P. Gambino and G.F. Giudice,
                    CERN preprint CERN--TH/98--177, 1998,
                    {\sl hep-ph/9806308}.


\bibitem{BT} J. Binney and S. Tremaine,
	{\em Galactic Dynamics} (Princeton Univ. Press, 1987).

\bibitem{nfw1} J.F. Navarro, C.S. Frenk and S.D.M. White, Ap. J. 462 (1996) 563.

\bibitem{nfw2} J.F. Navarro, C.S. Frenk and S.D.M. White, Ap. J. 490 (1997) 493.


\bibitem{kkbp} A.V. Kravtsov, A.A. Klypin, J.S. Bullock and J.R. Primack, 
Astrophys. J. 502 (1998) 48.

\bibitem{moore} B. Moore, T. Quinn, F. Governato, J. Stadel and G. Lake,
astro-ph/9903164.

\bibitem{turner_rhol} E. Gates, G. Gyuk and M.S. Turner,
                     Phys. Rev. Lett. 74 (1995) 3724;
                     Phys. Rev. D 53 (1996) 4238;
E. Gates, G. Gyuk and M.S. Turner,
  Astrophys. J. Lett. 449  (1995) L123;
E. Gates, G. Gyuk and M.S. Turner,
  {\sl astro-ph/9704253}, {\em Proceedings of the 18th Texas Symposium on
      Relativistic Astrophysics}, edited by A. Olinto, J. Frieman and 
      D. Schramm (World Scientific, to appear).

\bibitem{koch} C.S. Kochanek, Ap. J. 457 (1996) 228.

\bibitem{leonard} P.J.T. Leonard and S. Tremaine, Ap. J. 353 (1990) 486.

\bibitem{cud} K.M. Cudworth, Astron. J. 99 (1990) 590.

\bibitem{evans}  N.W. Evans, Mont. Not. R. Astr. Soc. 260
(1993) 191.

\bibitem{fv_rot} F. Donato, N. Fornengo and S. Scopel,
Astropart. Phys. 9 (1998) 247; M. Kamionkowsky and A. Kinkhabwala,
Phys. Rev. D 57 (1998) 3256.

\bibitem{witten} M.W. Goodman and E. Witten, Phys. Rev. D 31 (1985) 3059.


\bibitem{taup_direct} For an extensive and recent review
of the different experimental efforts involved in
direct detection, as well as indirect detection, see the Proceedings of 
TAUP 97, Nucl. Phys. B (Proc. Suppl.) 70
(edited by A. Bottino, A. Di Credico, P. Monacelli),
January 1999.

\bibitem{bdmsbi} A. Bottino, F. Donato, G. Mignola, S. Scopel, P. Belli and 
A. Incicchitti, Phys. Lett. B 402 (1997) 113.

\bibitem{DAMA_uplim} R. Bernabei et al., Phys. Lett. B389 (1996) 757.

\bibitem{noi_diretta} A.Bottino, V.de Alfaro, N.Fornengo, G.Mignola, S.    
   Scopel, C.Bacci et al., Phys. Lett. B259 (1992) 330;
A.Bottino, V.de Alfaro, N.Fornengo, G.Mignola and
  S. Scopel, Astropart. Phys. 2 (1994) 77;


\bibitem{diretta_theo}
M. Drees and M.M. Nojiri, Phys. Rev. D48 (1993) 3483;
P. Gondolo and L. Bergstr\"om, Nucl. Phys. (Proc.Suppl.) 48 (1996) 53;
R. Arnowitt and Pran Nath, Phys. Lett. B437 (1998) 344;
V. A. Bednyakov and H. V. Klapdor-Kleingrothaus, Phys.Rev. D 59 (1999) 023514;
T. Falk, A. Ferstl and K. A. Olive, Phys. Rev. D 59 (1999) 055009.


\bibitem{note_damour} 
In Ref. \cite{dk} it has been suggested that there may exist a population
of WIMPs which were set into bound Earth--crossing solar orbits by a
peripheral scattering of galactic WIMPs on the surface of the Sun, with
subsequent gravitational perturbations by planets. These WIMPs would have
a dispersion velocity smaller (roughly, by a factor of 10)  than the usual
one of galactic WIMPs and would enhance direct detection rates at very low
nuclear recoils (of order of KeV). Estimates of the expected effects are
provided in Ref. \cite{dk}. Using these results it turns out that the
upper limits on $\xi \sigma^{\mathrm nucleon}_{\mathrm scalar}$ of 
Ref. \cite{DAMA_uplim} already set
bounds on the model parameters which are somewhat more stringent than the
ones indicated in Ref. \cite{dk} (see also J.I. Collar,
Phys. Rev. D59 (1999) 063514 for a similar conclusion). For instance, 
for WIMPs of this population one would have
$m_{\chi} \gsim$ 200 GeV, unless $\rho_l \lsim $ 0.2 GeV cm$^{-3}$.


\bibitem{dk} T. Damour and L.M. Krauss, Phys. Rev. Lett. 81 (1998) 5726
and  Phys. Rev. D 59 (1999) 063509.



\bibitem{ann_mod_th} A.K. Drukier, K. Freese and D.N. Spergel, 
                        Phys. Rev. D 33 (1986) 3495;
                     K. Freese, J. Frieman and A. Gould, 
                        Phys. Rev. D 37 (1988) 3388.


\bibitem{DAMA_longrep} R. Bernabei et al., University of Rome Report No. 
                     ROM2F/98/27 (submitted for publication).

\bibitem{DAMA} R. Bernabei et al., Phys. Lett. B424 (1998) 195;
R. Bernabei et al., University of Rome Report No. 
 			ROM2F/98/34 and INFN Report No. INFN/AE-98/20 
                        (to appear in Phys. Lett. B).


\bibitem{bellietal} P. Belli, R. Bernabei, A. Bottino, F. Donato, N. Fornengo,
D. Prosperi and S. Scopel, {\sl hep--ph/9903501}. 

\bibitem{noi_DAMA} A. Bottino, F. Donato, N. Fornengo and S. Scopel, 
    Phys. Lett. B 423 (1998) 109;
 A. Bottino, F. Donato, N. Fornengo and S. Scopel, 
{\sl hep-ph/9710295}, Torino University Report No. DFTT 61/97
 (unpublished);
 A. Bottino, F. Donato, N. Fornengo and S. Scopel, 
{\sl hep-ph/9808456}, Torino University Report No. DFTT 41/98
 (to appear in Phys. Rev. D);
 A. Bottino, F. Donato, N. Fornengo and S. Scopel, 
{\sl hep-ph/9808459}, Torino University Report No. DFTT 48/98
 (to appear in Phys. Rev. D);
 A. Bottino, F. Donato, N. Fornengo and S. Scopel, 
                Astropart. Phys. 10 (1999) 203.

\bibitem{an} R. Arnowitt and P. Nath, {\sl hep-ph/9902237}. 

\bibitem{ritz-seckel} S. Ritz and D. Seckel, Nucl. Phys. B304 (1988) 877.

\bibitem{noi_nuflux} A.Bottino, N.Fornengo, G.Mignola, L.Moscoso,
Astropart. Phys. 3 (1995) 65;
A.Bottino, V.de Alfaro, N.Fornengo, A.Morales,       
   J.Puimedon and S.Scopel, Mod. Phys. Lett. A7 (1992) 733;
A.Bottino, V.de Alfaro, N.Fornengo, G.Mignola and
M .Pignone, Phys. Lett. B265 (1991) 57.

\bibitem{altri_nuflux} 
L. Bergstr\"om, J. Edsj\"o and P. Gondolo, Phys. Rev. D 58 (1998) 103519;
L. Bergstr\"om, J. Edsj\"o, M. Kamionkowski, Astropart. Phys. 7 (1997) 147;
F. Halzen, {\sl hep--ph/9506304}.

\bibitem{MACRO} The MACRO Collaboration, M. Ambrosio et al.,
{\sl hep--ex/9812020}.


\bibitem{noi_pbar}
A. Bottino, F. Donato, N. Fornengo and P. Salati,
Phys. Rev. D 58 (1998) 123503;
A. Bottino, C. Favero, N. Fornengo and G. Mignola,
Astropart. Phys. 3 (1995) 77.


\bibitem{others_pbar} 
 L. Bergstr\"om, J. Edsj\"o and P. Ullio, {\sl astro-ph/9902012};
J.D. Wells, A. Moiseev, J.F. Ormes, {\sl hep--ph/9811325};
T. Mitsui, K. Maki and S. Orito, Phys. Lett. B389 (1996) 169;
P. Chardonnet, G. Mignola, P. Salati and R. Taillet,
Phys. Lett. B384 (1996) 161.


\bibitem{dbar} F. Donato, N. Fornengo and P. Salati,
Valencia University preprint FTUV/99-9 and IFIC/99-9.

\bibitem{secondary} More detailed evaluations of the secondary antiproton
spectrum at low energies are still in progress; for some recent 
advancements in this field see  L. Bergstr\"om, J. Edsj\"o and 
P. Ullio, {\sl astro-ph/9902012} and 
J.W. Bieber, R.A. Burger, R. Engel, T.K. Gaisser, S. Roesler,
T. Stanev, {\sl astro--ph/9903163}.


\bibitem{BESS97} S. Orito, International Conference on High Energy Physics, 
                Vancouver, 1998.

\bibitem{suede_positron} E.A. Baltz, J. Edsj\"o, Phys. Rev. D59 (1999) 023511.

\bibitem{moskalenko} L.V. Moskalenko and A. Strong, 
Astrophys. J. 493 (1998) 694.

\bibitem{HEAT} S.W. Barwick et al. (HEAT Collaboration),
Astrophys. J. 498 (1998) 779.


\bibitem{gamma} 
L. Bergstr\"om and J. Kaplan, Astropart. Phys. 2 (1994) 261
J. Silk and A. Stebbins, Astrophys. J. 411 (1993) 439;
M. Urban et al., Phys. Lett. B 293 (1992) 149;
H.U. Bengtsson, P. Salati and J. Silk, Phys. Rev. D 40 (1989) 3828.

\bibitem{gamma_bbm} V. Berezinsky, A. Bottino and G. Mignola, 
		    Phys. Lett. B325 (1994) 136.


\bibitem{line_bba}
V. Berezinsky, A. Bottino and V. de Alfaro, Phys. Lett. B274 (1992) 122


\bibitem{bub} L. Bergstr\"om, P. Ullio and J.H. Buckley, Astropart. Phys. 9
(1998) 137.

\bibitem{suede_clump} 
L. Bergstr\"om, J. Edsj\"o, P. Gondolo and P. Ullio, 
Phys. Rev. D59 (1999) 043506.

\bibitem{gamma_annihil}
P. Ullio and L. Bergstr\"om, Phys. Rev. D 57 (1998) 1962;
L. Bergstr\"om and P. Ullio, Nucl. Phys. B504 (1997) 27;
Z. Bern, P. Gondolo and M. Perelstein, Phys. Lett. B 411 (1997) 86.


\bibitem{EGRET} P. Sreekumar et al., Astrophys. J. 494 (1998) 523.


\end{thebibliography}
\end{document}